\documentclass[fleqn,usenatbib]{mnras}

\RequirePackage{snapshot}

\usepackage{newtxtext,newtxmath}

\usepackage[T1]{fontenc}

\DeclareRobustCommand{\VAN}[3]{#2}
\let\VANthebibliography\thebibliography
\def\thebibliography{\DeclareRobustCommand{\VAN}[3]{##3}\VANthebibliography}

\usepackage{graphicx}	
\usepackage{amsmath}	

\newcommand{\vc}[1]{{\mathbf{#1}}}

\title[Precipitation in the CGM]{Precipitation possible:  turbulence-driven thermal instability with constrained entropy profiles}

\author[B.D. Wibking et al.]{
Benjamin D. Wibking,$^{1}$\thanks{E-mail: wibkingb@msu.edu (BDW)}
G. Mark Voit,$^{1}$
and Brian W. O'Shea$^{1,2,3}$
\\
$^{1}$Department of Physics and Astronomy, Michigan State University, 567 Wilson Road, East Lansing, MI 48824, USA\\
$^{2}$Department of Computational Mathematics, Science, and Engineering, Michigan State University, 428 South Shaw Lane, East Lansing, MI 48824, USA\\
$^{3}$Facility for Rare Isotope Beams, Michigan State University, 640 South Shaw Lane, East Lansing, MI 48824, USA\\
}

\date{Accepted XXX. Received YYY; in original form ZZZ}

\pubyear{2024}

\begin{document}
\label{firstpage}
\pagerange{\pageref{firstpage}--\pageref{lastpage}}
\maketitle

\begin{abstract}
Precipitation of cold gas due to thermal instability in  both galaxy clusters and the circumgalactic medium may regulate AGN feedback. We investigate thermal instability in idealized simulations of the circumgalactic medium with a parameter study of over 600 three-dimensional hydrodynamic simulations of stratified turbulence with cooling, each evolved for 10 Gyr. The entropy profiles are maintained in a steady state via an idealized `thermostat' process, consistent with galaxy cluster entropy profiles. In the presence of external turbulent driving, we find cold gas precipitates, with a strong dependence whether the turbulent driving mechanism is solenoidal, compressive, or purely vertical.
In the purely-vertical turbulent driving regime, we find that significant cold gas may form when the cooling time to free-fall
time $t_{\rm cool} / t_{\text{ff}} \lesssim 5$. Our simulations with a ratio of $t_{\rm cool} / t_{\text{ff}} \sim 10$ do not precipitate under any circumstances, perhaps because the thermostat mechanism we use maintains a significant non-zero entropy gradient.
\end{abstract}




\section{Introduction}

Galactic atmospheres have been a topic of astrophysical study ever since \citet{Spitzer_1956} proposed that the spiral arms of our Galaxy may form from cold clouds condensing out of a `Galactic corona.'
In the intervening decades, condensation via thermal instability (e.g., \citealt{Field_1965}) has been studied by many authors.
This paper focuses on thermal instability in \textit{stratified}\footnote{By ``stratified,'' we mean an atmosphere with a specific entropy gradient parallel to the gravitational potential gradient, in which localized vertical displacements excite internal gravity waves.} galactic atmospheres (e.g., \citealt{Defouw_1970,Balbus_1989}) because it has been proposed as a mechanism for self-regulation of kiloparsec-scale accretion flows that feed supermassive black holes in massive galaxies \cite[e.g.,][]{McCourt_2012,Sharma_2012}, perhaps with external sources of turbulence playing a critical role in riving the instability to non-linear amplitudes \citep[e.g.,][]{Gaspari_2013,Voit_2018}.

Observational evidence for the importance of thermal instability in stratified atmospheres has come from radial profiles of the cooling time to free-fall time ratio $t_{\rm cool}/t_{\text{ff}}$ in the cores of galaxy clusters \citep{Cavagnolo_2009,Voit_2015Natur.519..203V,Donahue_2022}. 
Values of $t_{\rm cool}/t_{\text{ff}}$ less than $\sim 30$ strongly correlate with the presence of cold clouds in cluster cores, and $t_{\rm cool}/t_{\text{ff}}$ is rarely observed to drop below $\sim 10$.
During the past decade, many theorists have been trying to explain the significance of this lower limit \citep[e.g.,][]{McCourt_2012, Sharma_2012, Gaspari_2013, Meece_2015, Choudhury_2016, Voit_2017, Gaspari_2018, Choudhury_2019a,Mohapatra_2019,Voit_2021,Mohapatra_2022,Mohapatra_2023,Mohapatra_2024}.

Here we present a systematic exploration of the relationship between thermal instability and turbulence in stratified galactic atmospheres, consisting of over 600 simulations of turbulence-driven thermal instability under conditions similar to the intra-cluster medium (ICM) in galaxy cluster cores.
We adopt a plane-parallel geometry and study idealized exponential atmospheres, for reasons explained later, and explore how an atmosphere's $t_{\rm cool}/t_{\text{ff}}$ ratio affects its susceptibility to precipitation.
In this context, ``precipitation'' refers to clouds that condense out of a hot galactic atmosphere, achieve density contrasts of non-linear amplitude, and rain down toward the center of the atmosphere's potential well.

A lower limit of $t_{\rm cool}/t_{\rm ff} \sim 1$ in galaxy cluster cores would be unsurprising, because stratified atmospheres with smaller $t_{\rm cool}/t_{\text{ff}}$ ratios are well known to be unstable to precipitation \citep[e.g.,][]{Defouw_1970,Balbus_1989,Binney_2009MNRAS.397.1804B,
McCourt_2012}.
However, buoyancy tends to suppress precipitation in stratified atmospheres with $t_{\rm cool}/t_{\text{ff}} \gg 1$ \citep{Cowie_1980,Nulsen_1986}, raising questions about whether thermal instability is really the origin of the $t_{\rm cool}/t_{\text{ff}} \gtrsim 10$ lower limit observed in galaxy cluster cores \citep[e.g.,][]{McNamara_2016}.
A stratified atmosphere with $t_{\rm cool}/t_{\text{ff}} \gg 1$ may still be thermally unstable, but damping of the density perturbations that develop prevents them from reaching non-linear amplitudes because of a mechanism that \citet{Voit_2017} called \textit{buoyancy damping} \citep[see also][]{McCourt_2012}: thermal pumping in such an atmosphere excites internal gravity waves that saturate with a density contrast $|\delta \rho / \rho| \sim t_{\rm ff} / t_{\rm cool}$ when energy losses via non-linear mode coupling balance energy input from thermal pumping.

However, turbulence can promote precipitation in atmospheres with $t_{\rm cool}/t_{\text{ff}} > 1$ if it can successfully interfere with buoyancy damping.
\citet{Gaspari_2013} showed with numerical simulations that driving of subsonic turbulence can promote precipitation in galactic atmospheres with $t_{\rm cool}/t_{\text{ff}} \approx 10$.
\citet{Voit_2018} extended that finding, using a heuristic analytical model to show how the amount of turbulence needed to induce precipitation might be related to an atmosphere's median $t_{\rm cool}/t_{\text{ff}}$ ratio. 
Essentially, the median value of $t_{\rm cool}/t_{\text{ff}}$ determines an atmosphere's \textit{susceptibility} to precipitation, because precipitation does not happen unless there are regions within the atmosphere where the \textit{local} value becomes $t_{\rm cool}/t_{\text{ff}} \lesssim 1$.
Whether or not precipitation actually occurs therefore depends on how the width of the distribution of $t_{\rm cool}/t_{\text{ff}}$ (i.e., its dispersion) compares with the median value \citep[see, e.g.,][]{Voit_2021}, and driving of turbulence helps to increase that dispersion.

Previous work on turbulence-driven precipitation in stratified atmospheres has explored the relevant parameter space only sparsely, with fewer than 20 simulations (e.g., \citealt{Mohapatra_2023}). This paper more densely samples that parameter space in order to reveal the precise boundaries between precipitating and non-precipitating regimes. We test three different modes of turbulent driving: solenoidal, compressive, and vertical. Solenoidal and compressive driving are common in the turbulence literature, but vertical driving is unusual since most simulations that incorporate turbulent driving do not have a gravitational potential or stratified medium, and thus have no geometric asymmetry.  We use it here to approximate the vertical motions induced by energy released as gas accretes onto a central supermassive black hole, also known as active galactic nucleus (AGN) feedback.

Another unique feature of the simulation suite is the thermostat mechanism we use to keep the background atmosphere steady. In most previous work on this topic the background atmosphere is allowed to evolve, sometimes making it unclear whether precipitation, if it occurs, is mostly a consequence of the atmosphere's \textit{initial} configuration or its \textit{later} configuration, when precipitation starts to happen. Our new thermostat mechanism minimizes that ambiguity by keeping the atmosphere's background configuration nearly constant.

The paper is organized as follows. Section \ref{section:methods} describes our numerical methods and the physics modeled in our simulations. Section \ref{section:results} describes the outcomes of our simulations with respect to their precipitation properties. Section \ref{section:discussion} discusses our results in the context of previous studies of precipitation. Section \ref{section:conclusion} outlines our conclusions and proposes some follow-up opportunities.

\section{Methods}
\label{section:methods}

We simulate idealized circumgalactic atmospheres in three-dimensional Cartesian geometry using the \textsc{AthenaPK} magnetohydrodynamics code, implemented using the \textsc{Parthenon} AMR framework \citep{Grete_2022}. Although \textsc{AthenaPK} is capable of simulating magnetic fields, we do not use that capability here.

The governing equations are those of inviscid hydrodynamics with source terms:
\begin{align}
    \frac{\partial \rho}{\partial t} + \nabla \cdot (\rho \vc{v})                              & = 0 \, ,       \\
    \frac{\partial (\rho \vc{v})}{\partial t} + \nabla \cdot (\rho \vc{v} \vc{v} + \mathsf{P}) & = \rho \vc{g} + \rho \vc{a} \, ,  \\
    \frac{\partial E}{\partial t} + \nabla \cdot \left[(E + \mathsf{P})\vc{v}\right]           & = \vc{v} \cdot (\rho \vc{g} + \rho \vc{a}) + \mathcal{H} - \mathcal{C} \, ,
\end{align}
where $\rho$ is the gas density, $\vc{v}$ is the fluid velocity, $E$ is the total energy density (kinetic and thermal),
$\mathsf{P}$ is the gas pressure tensor, $g$ is the gravitational acceleration, $a$ is the external turbulent acceleration, $\mathcal{H}$ is the heating rate per unit volume, and $\mathcal{C}$ represents cooling per unit volume.

The hydrodynamics is solved with the second order strong-stability-preserving Runge-Kutta (RK2-SSP) time integrator \citep{Shu_1989} combined with the piecewise-parabolic method (PPM; \citealt{Colella_1984}) for the reconstruction  of the primitive variables (density, velocity, and pressure), with the extrema-preserving modification of \cite{McCorquodale_2011}. A low-dissipation variant of the Harten-Lax-van Leer contact (HLLC) Riemann solver is used to compute the fluxes between cells \citep{Minoshima_2021}.\footnote{Standard Riemann solvers do \emph{not} produce the correct fluxes in the incompressible limit (i.e., $\mathcal{M} \rightarrow 0$) due to pressure fluctuations across a jump discontinuity scaling as $\mathcal{O}(\mathcal{M})$ rather than having the incompressible  scaling $\mathcal{O}(\mathcal{M}^2)$ \citep{Guillard_2004,Thornber_2008}.}

The gravitational source terms are handled using the well-balanced reconstruction method of \cite{Kappeli_2014}. This method subtracts the hydrostatic pressure before performing the reconstruction of the pressure variable and then restores the hydrostatic contribution to the pressure at the interfaces. This procedure requires an assumption about the nature of hydrostatic equilibrium within each simulation cell. We assume that each cell is isothermal, consistent with our initial conditions. We find that well-balancing is essential in order to avoid numerical artifacts at low Mach numbers ($\mathcal{M} \lesssim 0.1$).

The cooling term $\mathcal{C}$ is intended to represent radiative cooling in the circumgalactic medium. We choose the form
\begin{align}
    \mathcal{C} = n_{\rm H}^2 \Lambda_0
\end{align}
where $n_{\rm H}$ is the hydrogen number density and $\Lambda_0 = 10^{-22}$ in cgs units, approximately the value for solar metallicity gas at $10^6$ K. However, the precise value of the cooling rate is irrelevant to our calculations, since we always scale the density profile to obtain the desired $t_{\rm cool}/t_{\text{ff}}$ profile for a fixed gravitational potential. To make our results independent of the temperature normalization, the cooling rate is chosen to be independent of temperature. Physically, this corresponds to the minimum $\Lambda(T)$ near $10^7$ K between metal line cooling and free-free cooling for gas in collisional ionization equilibrium (CIE). We do not include any physical heating processes in our simulations (but see section \ref{section:thermostat} for how we maintain the entropy profile).

The cooling time $t_{\text{cool}}$ is
\begin{align}
t_{\text{cool}} = \frac{3 \, k_B T}{2 \, n \, \Lambda} \frac{1}{(X \mu)^2} \, ,
\end{align}
where $X = 1 - Y$ is the hydrogen mass fraction, $Y = 0.25$ is the helium mass fraction, and the mean mass per particle in units of the proton mass $m_p$ is
\begin{align}
\mu = \left( \frac{3}{4} \, Y + 2 \, (1 - X) \right)^{-1} \approx 0.6 \, .
\end{align}
Since the cooling time depends on the normalization convention for the cooling curve $\Lambda$, it is very important to correctly normalize the cooling time, as a factor of 5 difference may result from omitting the $(X \mu)^{-2}$ factor. This factor appears because we define the cooling time in terms of $n$, whereas $\Lambda$ is defined with respect to $n_{\rm H}$.

\subsection{Initial conditions}

Our simulations are vertically stratified (i.e., along the $z$-coordinate direction). While real galaxy clusters are spherical, we adopt a Cartesian gravitational potential both for simplicity and in order to enable direct comparison with previous work (e.g., \citealt{McCourt_2012,Choudhury_2016}). We note that \cite{Choudhury_2016} have found similar global eigenmodes between Cartesian and spherical geometry for stratified atmospheres. Reflecting boundary conditions are adopted in the vertical direction in order to ensure global mass conservation in our simulations. Reflecting vertical boundaries also ensure that the atmosphere is numerically quiet for a hydrostatic profile (M. Zingale, private communication). In the horizontal direction, we adopt periodic boundary conditions, consistent with both the usual linear analysis of thermal instability and prior simulation work (e.g., \citealt{McCourt_2012}). Our simulations consist of a box of size $100$ kpc $\times$ $100$ kpc $\times$ $200$ kpc discretized with a grid size of $128 \times 128 \times 256$ cells.

We assume that the background atmospheric temperature is constant, with its value $T_0$ related to the gravitational acceleration $\vc{g} = - \text{sgn}(z) \, g(z) \, \hat z$ (where $\text{sgn}$ is the sign function) via
\begin{align}
g(z) \equiv \frac{k T_0}{\mu m_p} \, z_0^{-1} \, \tanh^2 \left(\frac{|z|}{h}\right) \, ,
\end{align}
in which $z_0$ is the pressure scale height, $m_p$ is the proton mass, and $h$ is a gravitational smoothing scale near the midplane. At a distance $|z|$ from the midplane, the `free-fall' timescale is
\begin{align}
t_{\text{ff}}(z) = \sqrt{\frac{2|z|}{g(z)}} \; .
\end{align}
Although this is not the true free-fall timescale, we note that this is the time it would take for a test particle to fall from a height $|z|$ to the midplane in the limit $h \rightarrow 0$. In our simulations, this is a useful timescale that can be defined via the local value of $g(z)$.
Following \cite{McCourt_2012}, we disable the cooling source term $\mathcal{C}$ and heating source term $\mathcal{H}$ within a distance $h$ of the vertical midplane of the simulation, since the entropy profile approaches a constant value within this region and allowing thermal instability to grow in this region would be unrepresentative of the global conditions implied by the mean value of ratio $t_{\text{cool}}/t_{\text{ff}}$.

We then compute the background density profile in hydrostatic equilibrium. For the number density profile $n(z)$, we have:
\begin{align}
    n(z) = n (z_0) \exp \left[ 1 - \frac {|z|} {z_0} 
        + \frac {h} {z_0} \left( \tanh \frac {|z|} {h} - \tanh \frac {z_0} {h} \right) \right] \, ,
\end{align}
where
\begin{align}
    n(z_0) = \frac {3(kT_0)^{3/2}} {2 \Lambda (X \mu)^2  (2 \mu m_p)^{1/2}}   
            \frac {\tanh (z_0/h)} {z_0} \left( \frac {t_{\rm cool}} {t_{\rm ff}} \right)_{z_0}^{-1} \, ,
\end{align}
and the ratio $t_{\rm cool}/t_{\text{ff}}$ is proportional to $|z/z_0|^{-1/2} \, n^{-1}(z)$.

\begin{figure*}
    \includegraphics[width=\textwidth]{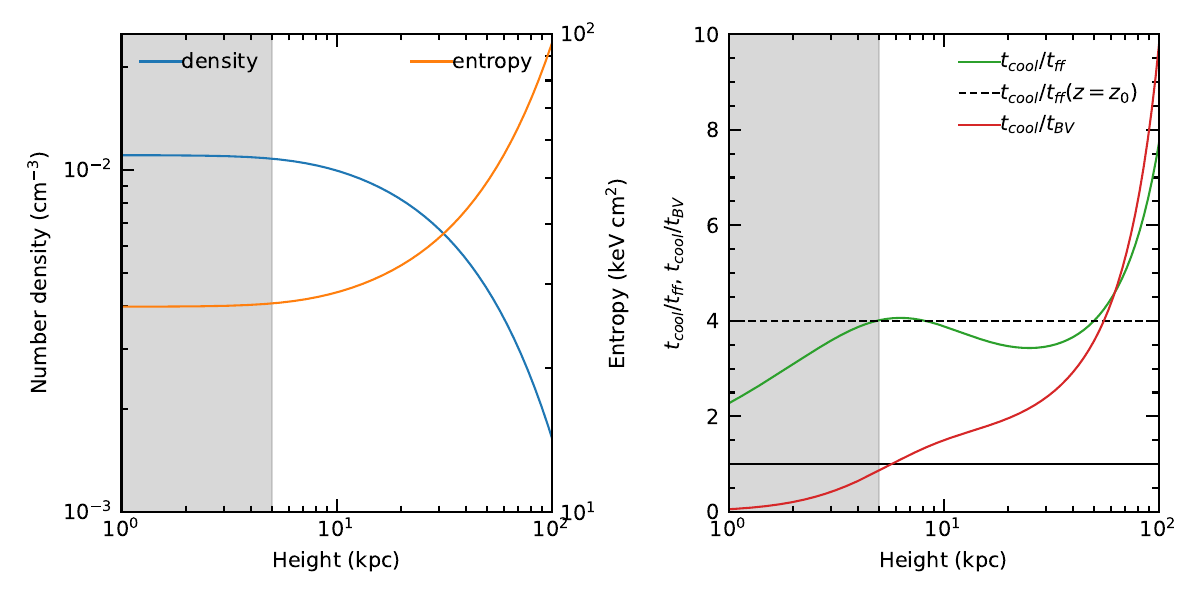}
    \caption{\emph{Left:} Density and entropy profiles as a function of height $|z|$ above the midplane for a fiducial cooling time to free-fall time ratio of $t_{\rm cool}/t_{\text{ff}} = 4$.
    The profiles approach a constant value below a height of $z \lesssim 10$ kpc because the gravitational acceleration is tapered to zero in order to avoid a discontinuous derivative of $\mathbf{g}$ across the midplane. In both panels, the shaded region marks the range of altitudes where heating and cooling are disabled. \emph{Right:} The profile of the $t_{\rm cool}/t_{\text{ff}}$ ratio and the profile of the ratio $t_{\rm cool}/t_{\text{BV}}$ relating the cooling time to the Brunt-V\"ais\"al\"a timescale.
    The $t_{\rm cool}/t_{\text{ff}}$ ratio remains similar to its value at the atmosphere's scale height $z_0 = 50 \, {\rm kpc}$ in the interval $5 \, {\rm kpc} \gtrsim |z| \gtrsim 50 \, {\rm kpc}$, whereas the ratio $t_{\rm cool}/t_{\text{BV}}$ approaches zero below one scale height. The dashed horizontal line indicates the value of $t_{\rm cool}/t_{\text{ff}}$ at the scale height $z_0$, whereas the solid horizontal line indicates a value of unity.}
    \label{fig:profiles}
\end{figure*}

We choose the pressure scale height $z_0$ to be half the distance from the midplane to the top of the box, and so $z_0 = 50$ kpc. This choice ensures that we resolve the pressure scale height with $10$-$20$ cells everywhere in the simulation, which is essential for maintaining hydrostatic equilibrium at very low Mach numbers \citep{Zingale_2002}. We adopt $h = 5$ kpc for the profile smoothing scale to ensure that $dg/dz$ remains continuous at the midplane. With these parameter values, we use the hydrostatic equilibrium condition to obtain the background profiles of density, entropy, and ratio of cooling time to free-fall time $t_{\rm cool}/t_{\text{ff}}$ shown in Figure \ref{fig:profiles}.

\subsection{Turbulent driving}
\label{subsection:driving}

For a subset of our simulations, we add momentum and kinetic energy to the gas via an acceleration source term in order to provide a source of external turbulent driving.
In particular, we use an Ornstein-Uhlenbeck process (e.g., \citealt{Eswaran_1988, Grete_2018}), where the Fourier components of the acceleration field
are sampled from a Gaussian at each wavenumber $k$. The power spectrum is chosen to be an downward-facing parabola centered on the wavenumber
corresponding to half of the size of the simulation box ($|k| = 2$ in units of inverse box length). This is the scale of the largest possible turbulent eddy, and is the scale at which we expect linear thermal instability to be least affected by buoyancy damping \citep{Voit_2017,Voit_2018}.

At each timestep, the real and imaginary components of each mode are advanced using the formal solution of the stochastic differential equation describing the Ornstein-Uhlenbeck process such that the impulses are correlated in time with a correlation timescale of $500$ Myr. This timescale is very nearly equal to the largest eddy turnover time in our simulation box, as is standard practice in many idealized simulations of turbulence (e.g., \citealt{Grete_2018}). Smaller eddies develop as turbulent kinetic energy cascades toward larger wavenumbers, which is then sustained in a quasi-steady state.

There are three driving configurations: solenoidal, compressive, and vertical. The solenoidal driving configuration consists of a Gaussian random vector field $\textbf{a}$ with the constraint that $\nabla \cdot \textbf{a} = 0$ (i.e., the field is divergence-free). The compressive driving configuration consists of a Gaussian random vector field $\vec{a}$ with the constraint that $\nabla \times \textbf{a} = 0$ (i.e., the field is curl-free). The vertical driving configuration consists of a Gaussian random vector field $\textbf{a}$ with the constraint that $a_x = 0$, $a_y = 0$, and $\partial a_z / \partial z = 0$ holds everywhere. This creates an acceleration field that perturbs gas in the $z$-direction with modulations in the $x$ and $y$ directions. The vertical configuration is motivated by the nature of AGN jets and galactic outflows, which provide a driving force that always acts in the direction opposing stratification (which is the vertical direction in our simulations), while maintaining the conventional practice of using a Gaussian random field for the turbulent driving acceleration field in idealized simulations of turbulence (e.g., \citealt{Eswaran_1988}).

\subsection{Thermostat}
\label{section:thermostat}

In order to study how thermal instability and precipitation depend on background conditions, those background conditions should be kept from evolving on a timescale comparable to the timescale for thermal instability to develop. A popular method for preserving the background profile has been to ensure that heating balances radiative cooling, on average, at each altitude\footnote{Sometimes called `magic heating'.} \citep{McCourt_2012}. At each time step, the radiative losses from each atmospheric layer are calculated, and an equivalent amount of thermal energy is added to the layer, usually via a heating rate per unit volume $\mathcal{H}(z)$ that depends only on altitude $z$. This heating method helpfully prevents the atmosphere from collapsing toward the origin on a cooling timescale, while allowing density perturbations within the atmosphere to grow on a cooling timescale.

However, we found while testing our simulations that this simple thermostat mechanism does not prevent the background conditions from changing when driven turbulence is added to the environment. First, dissipation of turbulence adds heat, and more vigorous driving adds even more heat. Second, driven turbulence promotes vertical diffusion of both kinetic and thermal energy, which can cause undesirable evolution of the background atmosphere. Turbulent diffusion in an isolated system would be expected to naturally flatten its entropy profile over time \citep{Wang_2023}, but here we seek to replicate and sustain an environment similar to the observed entropy profiles of galaxy clusters, which are typically power-law profiles with an inner core \citep{Cavagnolo_2009}.

Atmospheric entropy profiles require particularly close attention, because they determine whether or not an atmosphere is stable to convection, supports internal gravity waves, and suppresses precipitation through buoyancy damping \citep[see, e.g,][for a review]{Donahue_2022}. According to the classic Schwarzschild criterion, an atmosphere is convectively stable if its specific entropy increases with altitude \citep{Schwarzschild_1958ses..book.....S}. Figure \ref{fig:profiles} represents specific entropy in terms of the quantity $k T n_e^{-2/3}$, because the logarithm of $k T n_e^{-2/3}$ is proportional to the specific entropy of a monatomic ideal gas, with an adiabatic index $\gamma$ equal to 5/3.

Rather than attempting to construct a physical model to produce
a radial profile of entropy that is approximately constant in time, we instead add an artificial
`thermostat' term to the energy equation of the form
\begin{align}
    \mathcal{H} = - \frac{K_p}{t_{\text{cool}}(z)} \, \rho c_v \, \langle \Delta T \rangle_z \, ,
\end{align}
where $c_v$ is the specific heat at constant volume for a fully-ionized ideal gas of our chosen mean molecular weight (see Section \ref{section:methods}), ${\langle \Delta T \rangle}_z$ denotes the volume average of the temperature difference $\Delta T \equiv T(z) - T_0$ at height $z$, $K_p$ is a dimensionless prefactor for the constant of proportionality, and $t_{\text{cool}}(z)$ is the cooling time of the background profile at height $z$.
This yields a `thermostat' that continually pushes the atmosphere's background temperature profile toward the initial profile on a timescale proportional to the local cooling time. We have found that this method keeps the atmosphere's background configuration from evolving in the ways observed in previous work on thermal stability in turbulent atmospheres (e.g., \citealt{Mohapatra_2019,Mohapatra_2022,Mohapatra_2023,Mohapatra_2024}).

\subsection{Timescales}






Here we will introduce some timescales and dimensionless numbers describing our simulations that will helpful later, when we discuss our results.

\subsubsection{Buoyancy timescale}

Small-amplitude density perturbations in a stratified atmosphere oscillate at the Brunt-V\"ais\"al\"a period determined by buoyancy. When expressed in terms of a specific entropy gradient, the Brunt-V\"ais\"al\"a timescale (equal to the inverse of the Brunt-V\"ais\"al\"a oscillation frequency) is
\begin{align}
t_{\text{BV}} = \left( \frac {g} {\gamma} \frac {d} {dz} \ln Tn^{1-\gamma} \right)^{-1/2} ,
\end{align}
where $\gamma$ is the adiabatic index. Importantly, buoyancy damping becomes ineffective for $t_{\rm cool}/t_{\rm BV} \lesssim 1$ \citep[e.g.,][]{Voit_2017}, meaning that the flattening of the specific entropy gradient toward the center of our simulated atmospheres (see Figure \ref{fig:profiles}) may help to promote precipitation there.

To obtain a radial $t_{\rm BV}$ profile for the background atmosphere, recognize that $T = T_0$ is constant, meaning that $d \ln Tn^{1 -\gamma} / dz = - (\gamma - 1) \, d \ln P / dz = (\gamma - 1) g \rho / P$, and so
\begin{align}
t_{\text{BV}} (z) 
        = \frac {1} {\sqrt{\gamma - 1}} \frac {c_{\rm s}} {g(z)}     
        = \frac {\gamma} {\sqrt{\gamma - 1}} \frac {z_0} {c_{\rm s}} 
            \tanh^{-2} \left( \frac {|z|} {h} \right) \; \; ,    
\end{align}
where $c_{\rm s} \equiv \sqrt{\gamma kT_0 / \mu m_p}$ is the adiabatic sound speed at a temperature $T_0$. In our simulations, the ratio of $t_{\rm ff}$ to $t_{\rm BV}$ is therefore
\begin{align}
    \frac {t_{\rm ff}} {t_{\rm BV}} 
        = \sqrt{ \frac {2 (\gamma-1)}  {\gamma} \frac {|z|} {z_0} }
        \tanh \left( \frac {|z|} {h} \right)
        \label{eq:tff_tBV}
\end{align}
and $t_{\rm BV}$ is nearly independent of $z$ for $|z| > 10 \, {\rm kpc}$.

\subsubsection{Cooling timescale}

As discussed earlier, the cooling time $t_{\text{cool}}$ in our simulations depends on altitude according to:
\begin{align}
t_{\rm{cool}}(z) = & \, t_{\rm ff} (z_0) \, 
    \left(\frac{t_{\text{cool}}}{t_{\text{ff}}} \right)_{z_0} \, \nonumber \\
    & \: \times \exp \left[ - 1 + \frac {|z|} {z_0} 
        - \frac {h} {z_0} \left( \tanh \frac {|z|} {h} - \tanh \frac {z_0} {h} \right) \right].
\end{align}
It therefore increases nearly exponentially with height at $|z| > 10 \, {\rm kpc}$.

Using equation (\ref{eq:tff_tBV}) to convert from $t_{\rm cool}/t_{\text{ff}}$ to $t_{\rm cool}/t_{\text{BV}}$ at $|z| = z_0$ gives
\begin{align}
\left( \frac{t_{\text{cool}}}{t_{\text{BV}}} \right)_{z_0} &= 
\sqrt{\frac{2 (\gamma - 1)}{\gamma}} \, \left(\frac{t_{\text{cool}}}{t_{\text{ff}}} \right)_{z_0} \, \tanh \left( \frac {z_0} {h} \right), \\
&\approx 0.9 \, \left(\frac{t_{\text{cool}}}{t_{\text{ff}}} \right)_{z_0} \, ,
\end{align}
and so the ratios $t_{\rm cool}/t_{\text{ff}}$ and $t_{\rm cool}/t_{\text{BV}}$ are nearly equal at one scale height.

\subsubsection{Eddy turnover timescale}

Driving turbulence produces eddies that can inhibit precipitation by shredding incipient density perturbation faster than they can grow. In an atmosphere with typical velocity differences similar to $\delta v$, the eddy turnover timescale is $t_{\rm eddy} = L / \delta v$, where $L$ is the length scale associated with the drivers of turbulence.

\subsubsection{Dimensionless parameters}

From these timescales, we can define two dimensionless parameters: the \emph{Froude number} and the \emph{turbulent Damkohler number}.
The Froude number is the ratio of the Brunt-V\"ais\"al\"a period to the eddy turnover timescale:
\begin{align}
\text{Fr} &\equiv \frac{t_{\text{BV}}}{t_{\text{eddy}}} \approx \frac{\delta v}{L} \frac{\gamma}{\sqrt{\gamma - 1}} \, \frac {z_0} {c_s} = \frac{\gamma}{\sqrt{\gamma - 1}} \, \frac {z_0} {L} \mathcal{M}  \, \, ,
\end{align}
in which $\mathcal{M} = \delta v / c_{\rm s}$ is the turbulent Mach number.
In the limit $\text{Fr} \ll 1$, a stratified system experiences very fast buoyant restoring of perturbations, whereas in the limit $\text{Fr} \gg 1$, buoyancy acts on a timescale that is long compared to the rate at which turbulent eddies disturb the stratification of the fluid, meaning that departures from hydrostatic equilibrium may become significant. Observationally, galaxy clusters have $\text{Fr} \lesssim 1$ (although this depends on whether there is an entropy core on scales $\lesssim 10$ kpc; see \citealt{Wang_2023}).

For our simulations, the Froude number is nearly constant with height and scales with the ratio of pressure scale height to driving scale $z_0 / L$ and with the resulting Mach number $\mathcal{M}$ of the simulation. Since it depends on the Mach number of each simulation as a function of simulation parameters, we cannot precisely specify the Froude number of our simulations \emph{a priori}. However, we can compute its value as a function of Mach number for our simulation setup:
\begin{align}
    \text{Fr}
    &\approx 2.0 \, \mathcal{M}  \, ,
\end{align}
since the pressure scale height is equal to the driving scale in our simulations. For observationally-relevant Mach numbers $\mathcal{M} \approx 0.3$, we have $\text{Fr} \approx 0.6$. This is comparable to the observationally-inferred Froude number of galaxy clusters on scales of $10 \lesssim r \lesssim 100$ kpc (e.g., \citealt{Wang_2023}).

The turbulent Damköhler number $\text{Da}_t$ is the ratio of the eddy turnover timescale to the cooling timescale. In the limit $\text{Da}_t \ll 1$, turbulent motions are much faster than the cooling time, so cooling has little effect. In the opposite limit $\text{Da}_t \gg 1$, cooling is much faster than turbulent motions, so that cooling may be the dominant physical process determining the dynamics. For our simulations, at altitudes where $\tanh (|z|/h) \approx 1$, we have:
\begin{align}
\text{Da}_t \approx \frac{1}{\sqrt{2 \gamma}} \left( \frac{t_{\text{cool}}}{t_{\text{ff}}} \right)_{z_0}^{-1} \, 
\frac{L} {z_0} \, \mathcal{M}^{-1} \exp{ \left( 1 - \frac{z}{z_0} \right) } \, ,
\end{align}
which decreases exponentially with height. This number cannot be precisely specified \emph{a priori} for our simulations due to its dependence on the Mach number $\mathcal{M}$. However, we can give an order-of-magnitude estimate of its value at one scale height:
\begin{align}
    \text{Da}_t(z_0) \approx 0.55 \left( \frac{t_{\text{cool}}}{t_{\text{ff}}} \right)_{z_0}^{-1} \, \mathcal{M}^{-1} \, ,
\end{align}
where for observationally-relevant Mach numbers $\mathcal{M} \approx 0.3$, we have
\begin{align}
    \text{Da}_t(z_0) \approx 1.8 \left( \frac{t_{\text{cool}}}{t_{\text{ff}}} \right)_{z_0}^{-1} \, .
\end{align}
For sufficiently small values of $\text{Da}_t$, we expect turbulent mixing effects to become more important than cooling (although the exact value cannot be determined a priori).

\subsection{Parameter space}
The parameter space we explore is the Mach number--$t_{\rm cool}/t_{\text{ff}}$ plane. The range of $t_{\rm cool}/t_{\text{ff}}$ is chosen to be from 1 to 10, sampled in $\log_{10} t_{\rm cool}/t_{\text{ff}}$. This choice is motivated by the theoretical boundary of $t_{\rm cool}/t_{\text{ff}} = 1$ and by the observationally significant value $t_{\rm cool}/t_{\text{ff}} = 10$. At a ratio of 1 or below gas should always precipitate since classical buoyancy damping should be ineffective, but galaxy clusters with $t_{\rm cool}/t_{\text{ff}}\approx10$ looks like they may still be susceptible to precipitation (e.g., \citealt{Cavagnolo_2009,Voit_2015Natur.519..203V,Donahue_2022}).

The range of turbulent driving amplitude $A_{\text{drive}}$ is chosen from extensive numerical experiments with our turbulence driving scheme and initial conditions such that it produces turbulent Mach numbers $\mathcal{M}$ between $\mathcal{M} \sim 10^{-2}$ and $\mathcal{M} \sim 1$. These Mach numbers encompass the range of observed turbulent Mach numbers of galaxy clusters in the range
$0.2 \lesssim \mathcal{M} \lesssim 0.3$ (e.g., \citealt{Zhuravleva_2015,Hitomi_2016,Hitomi_2018,Zhuravleva_2023}).

The joint distribution of $\log_{10} t_{\rm cool}/t_{\text{ff}}$ and $A_{\text{drive}}$ is sampled using a Halton quasi-random sequence \citep{Halton_1960} using the \textsc{Chaospy} software package \citep{Feinberg_2015}. The use of a quasi-random sequence is motivated by the need to more efficiently sample the two-dimensional parameter space than is obtained by independent random samples.\footnote{Empirically, quasi-random sampling commonly requires $1-2$ orders of magnitude fewer samples when compared to simple random sampling in order to obtain a fixed level of accuracy for integrals over a given parameter space \citep{Berblinger_1991}.}

\section{Results}
\label{section:results}
\subsection{No external driving}

First we report results from simulations without driven turbulence. We initialize the simulations in hydrostatic equilibrium, and then add percent-level density fluctuations to seed thermal instability (for the simulations described in this subsection only). The resulting turbulence is then purely a result of thermal instability.

\begin{figure}
    \includegraphics[width=\columnwidth]{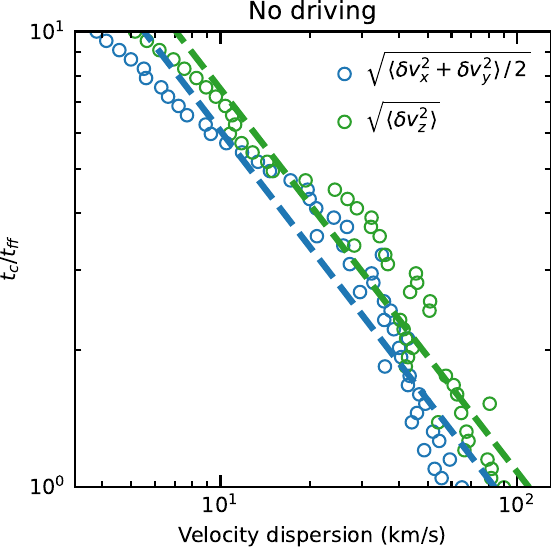}
    \caption{The rms velocity dispersion in vertical and horizontal directions for varying $t_{\rm cool}/t_{\text{ff}}$ for simulations without external turbulent driving. For each simulation, the velocity dispersions are measured at the simulation output at which the midplane cold gas mass fraction $f_{\text{cold}}$ is at its maximum value.}
    \label{fig:rms_dv_nodriving}
\end{figure}

Those results are shown in Figure \ref{fig:rms_dv_nodriving}. The velocity dispersion of the simulations is defined as the rms velocity $\sqrt{\langle (v(z) - \bar v(z))^2 \rangle}$  measured at a height $z = 10$ kpc (with all averages volume-weighted). The blue circles show the rms horizontal velocity dispersion $\sqrt{ \langle \delta v_x^2 + \delta v_y^2 \rangle / 2 }$. The green circles show the rms vertical velocity dispersion $\sqrt{\langle \delta v_z^2 \rangle}$. The dashed lines show the power-law fits to these points. The rms horizontal velocity dispersion is reasonably well fit by a power law, computed via ordinary least-squares (OLS) regression applied in log-log space:
\begin{align}
{\langle \delta v_{\text{xy}} \rangle}_{\text{rms}} = \left( \frac{t_{\rm cool}}{t_{\text{ff}}} \right)^{-1.18} \, 84.0 \, \text{km s}^{-1}  \, .
\end{align}
The rms vertical velocity dispersion is reasonably well fit by the power law:
\begin{align}
{\langle \delta v_{\text{z}} \rangle}_{\text{rms}} = \left( \frac{t_{\rm cool}}{t_{\text{ff}}} \right)^{-1.19} \, 110. \, \text{km s}^{-1} \, .
\end{align}
For reference, we note that the adiabatic sound speed of a fully-ionized H/He plasma (with helium mass fraction $Y = 0.25$) at our background temperature $T_0 = 10^{7}$ K is $484$ km s$^{-1}$, implying turbulent Mach numbers $\mathcal{M} \sim 0.2$.

We see that the horizontal velocity dispersions are systematically offset to the left compared to the vertical velocity dispersions, indicating that the average ratio of vertical to horizontal velocity dispersions is greater than unity.
Vertical velocity dispersions are greater because gravity responds to thermal instability by pulling the denser blobs down while the less density ones rise, so there is vertical forcing in addition to whatever motions thermal instability would cause in the absence of gravity.
We note that in these simulations, typical galaxy cluster velocity dispersions ($\sim 100$ km/s) are only obtained for ratios of $t_{\rm cool}/t_{\text{ff}} \lesssim 3$.
\\

\begin{figure}
    \includegraphics[width=\columnwidth]{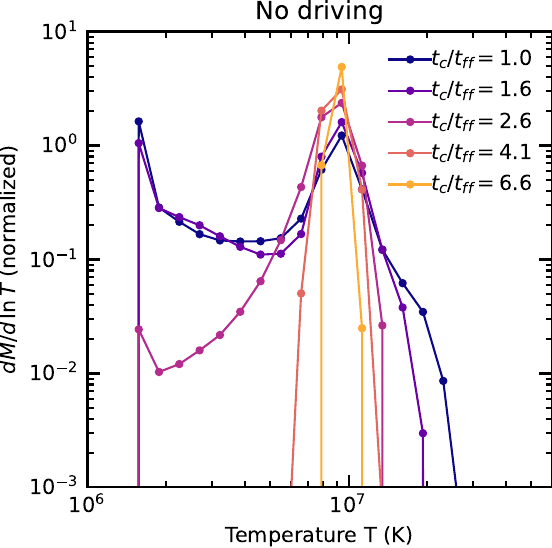}
    \caption{The probability distribution function by mass of temperature for varying $t_{\rm cool}/t_{\text{ff}}$.}
    \label{fig:pdf_temp_nodriving}
\end{figure}

In Figure \ref{fig:pdf_temp_nodriving}, we show the probability distribution function (PDF) of temperature weighted by mass in a subset of the simulations without turbulent driving. 
The distributions with $t_{\rm cool}/t_{\text{ff}} > 3$ are approximately log-normal and centered on the initial temperature of $10^7$ K. As $t_{\rm cool}/t_{\text{ff}}$ decreases, the width of the distributions increases. Below $t_{\rm cool}/t_{\text{ff}} \approx 3$, a second peak in the temperature appears just above the temperature floor of $10^6$ K, indicating that a significant amount of cold gas has precipitated. Qualitatively, these results are expected based on the standard analysis of precipitation in the absence of external sources of turbulence (e.g., \citealt{Voit_2021}; see also \citealt{Mohapatra_2023}).

\subsection{External driving}

\begin{figure*}
    \includegraphics[width=\textwidth]{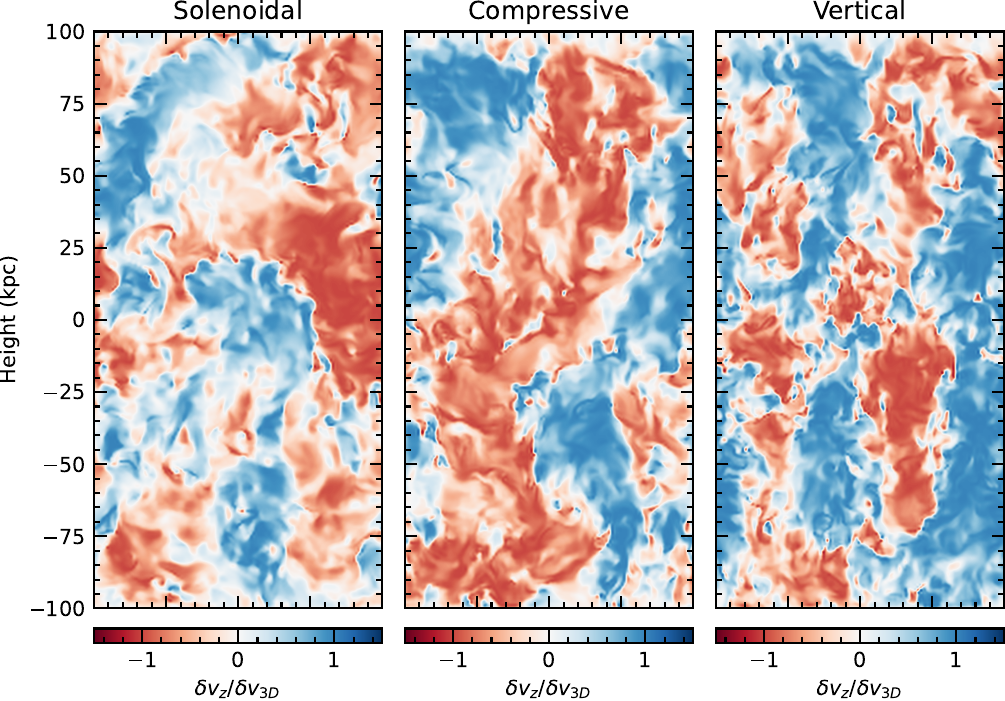}
    \caption{A vertical slice of the velocity anisotropy parameter $\delta v_z / \sqrt{\delta v_{3D}^2}$ through three simulation boxes for solenoidal, compressive, and vertical driving simulation with similar Mach numbers $\mathcal{M} \approx 0.25$ and $t_{\rm cool}/t_{\text{ff}} \approx 3.8$ for the simulation output containing the maximum midplane ($\pm 10$ kpc) cold gas fraction $f_{\text{cold}}$. The temperature PDF of these simulations are also shown in Figure 9.}
    \label{fig:slice_vel}
\end{figure*}

We now turn to our simulations that included external turbulent driving. Three distinct
types of turbulent driving are used (as described previously in Section \ref{subsection:driving}): solenoidal, compressive, and vertical. Each type of driving leaves its imprint in the velocity field of the simulation in a unique way. The leftmost panel of Figure \ref{fig:slice_vel} shows the
imprint on the vertical velocity field left by solenoidal driving, wherein we see flocculant
patches of red and blue, indicating negative and positive vertical-to-total velocity fluctuations
$\delta v_z / \sqrt{\delta v^2}$, respectively. In the middle panel, showing a compressively-driven
simulation, we see larger coherent patches of positive and negative vertical-to-total velocity
fluctuations, respectively. In the rightmost panel, showing a vertically-driven simulation,
there are highly coherent patches of positive or negative vertical-to-total fluctuations in the velocity field with near-symmetry about the horizontal $z=0$ plane.

\begin{figure*}
    \includegraphics[width=\textwidth]{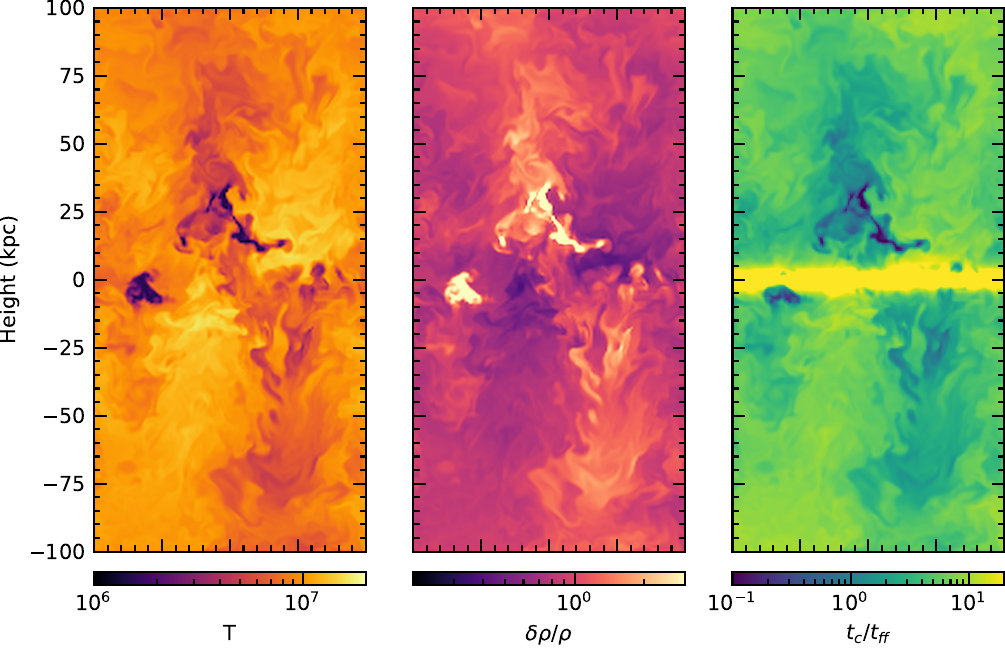}
    \caption{A vertical slice of temperature, density contrast, and ratio of cooling time to free-fall time through the simulation box for a vertical driving simulation with Mach number $\mathcal{M} = 0.24$ and $t_{\rm cool}/t_{\text{ff}} = 3.75$ for the simulation output containing the maximum midplane ($\pm 10$ kpc) cold gas fraction $f_{\text{cold}}$. The temperature PDF of this simulation is also shown in Figure 9.}
    \label{fig:slice}
\end{figure*}

A representative simulation slice for vertical driving is shown in Figure \ref{fig:slice}. This simulation has a Mach number $\mathcal{M} \approx 0.24$ and a ratio of cooling time to free-fall time $t_{\rm cool}/t_{\text{ff}} = 3.75$. This slice is chosen as the vertical slice that contains the maximum midplane cold gas fraction $f_{\text{cold}}$, defined as the ratio of cold gas to total gas mass within $\pm 10$ kpc of the midplane (see, e.g., \citealt{Choudhury_2016,Choudhury_2019b}). We see a significant amount of condensed cold gas near the $z=0$ midplane, reflecting the structures visible in all three panels (from left to right: temperature, local density contrast, and ratio of cooling time to free-fall time). Vertical plumes of relatively colder gas extending both above and below the midplane are visible in all three panels. These features are all typical for simulations with a significant midplane cold gas fraction.

\subsubsection{Solenoidal driving}

\begin{figure*}
    \includegraphics[width=\textwidth]{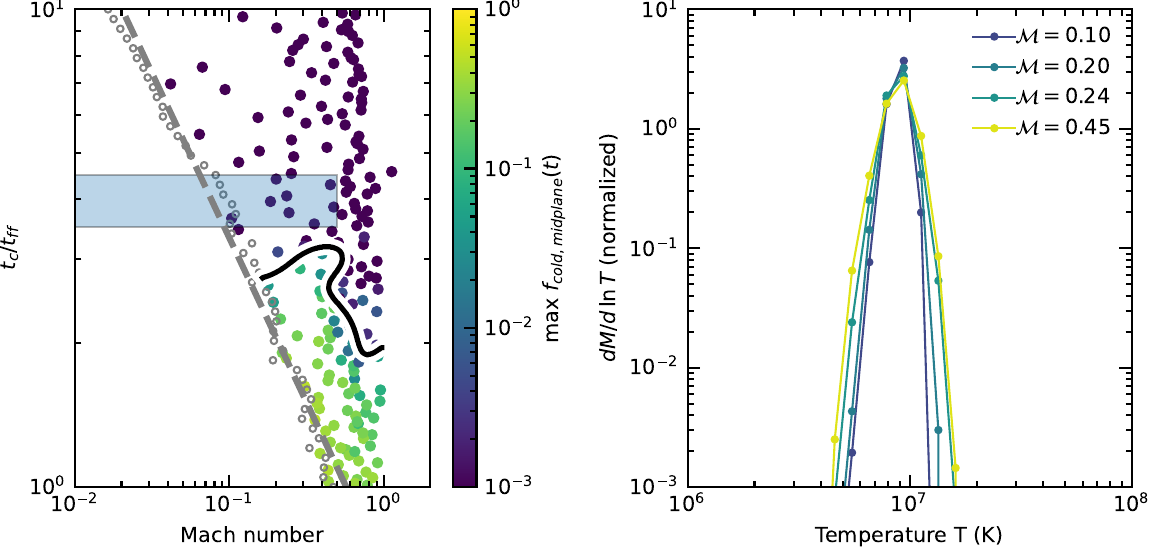}
    \caption{\emph{Left:} For solenoidal driving, the maximum fraction of cold gas in the midplane $f_{\text{cold}}$ as a function of cooling-to-freefall-time ratio and near-midplane rms sonic Mach number $\mathcal{M}$. The filled circles show simulations with turbulent driving, while the unfilled circles represent simulations without turbulent driving. The shaded box shows the regime from which the profiles shown in the right panel are drawn. The dashed gray line shows the best-fit power-law for the simulations without turbulent driving. The solid black line shows the boundary between precipitating and non-precipitating simulations with turbulent driving constructed using a support vector machine (SVM) classifier, where precipitating is defined as having $f_{\text{cold}} > 10^{-2}$. \emph{Right:} The probability distribution function of temperature for varying amplitudes of solenoidal driving for simulations where $3.5 < t_{\rm cool}/t_{\text{ff}} < 4.5$. The amplitude of turbulent driving is shown in terms of the resulting turbulent rms Mach number.}
    \label{fig:solenoidal_driving}
\end{figure*}

Figure \ref{fig:solenoidal_driving} shows the outcomes of simulations with solenoidal driving. They come from a parameter study of 200 simulations run with solenoidal driving over a range of input $t_{\rm cool}/t_{\text{ff}}$ ratio and input turbulent driving amplitude. We sample the properties of each simulation at intervals of 500 Myr. From these samples, cold gas fraction and sonic Mach number are measured at the epoch at which the cold gas fraction is the highest. The dashed gray line in this Figure is a power-law fit to the simulations without driving, included as a reference representing the baseline scenario without external sources of turbulence. Without driving, the ratio of cooling time to free-fall time must be $\lesssim 3$ in order to produce any cold gas in the midplane.

The solid black line shows the boundary between precipitating and non-precipitating simulations, found by fitting a support vector machine to the simulation points with a threshold value of $f_{\text{cold}} = 10^{-2}$. The cold gas fraction $f_{\text{cold}}$ is defined as the ratio of cold gas mass to total gas mass within $\pm 10$ kpc of the midplane. The rms sonic Mach number is defined as $\sqrt{\langle (v/c_s - \langle v/c_s \rangle_z)^2 \rangle_z}$, where the averages $\langle \, . \, \rangle_z$ are area-weighted at height $z = 10$ kpc.

The shaded gray box indicates the region from which simulations are sampled in order to further examine the probability distribution function (PDF) of temperatures, shown in the left panel of Figure \ref{fig:solenoidal_driving}. In Figure \ref{fig:solenoidal_driving} (right panel), we show the PDFs corresponding to the simulations selected from the gray box at the epoch of maximum cold gas fraction. These simulations are selected to satisfy the criteria: $t_{\rm cool}/t_{\text{ff}}$ between 3.5 and 4.5, and sonic Mach numbers $\mathcal{M} < 0.5$. All of the temperature PDFs are unimodal and approximately log-normal, centered near a temperature of $10^7$ K. We observe that the width of these distributions increases monotonically with increasing Mach number, as expected from the stochastic precipitation model of \cite{Voit_2018,Voit_2021}. No simulations in this regime produce cold gas near the temperature floor of $10^6$ K.

\subsubsection{Compressive driving}

\begin{figure*}
    \includegraphics[width=\textwidth]{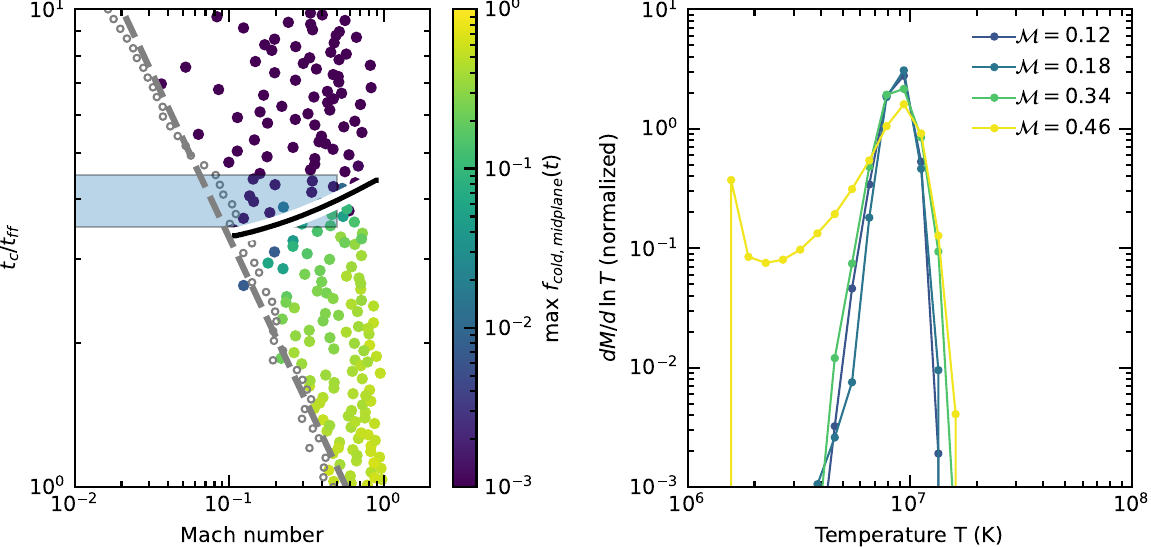}
    \caption{\emph{Left:} For compressive driving, the maximum fraction of cold gas in the midplane $f_{\text{cold}}$ as a function of cooling-to-freefall-time ratio and near-midplane rms sonic Mach number $\mathcal{M}$. The filled circles show simulations with turbulent driving, while the unfilled circles represent simulations without turbulent driving. The shaded box shows the regime from which the profiles shown in the right panel are drawn. The dashed gray line shows the best-fit power-law for the simulations without turbulent driving. The solid black line shows the boundary between precipitating and non-precipitating simulations with turbulent driving constructed using a support vector machine (SVM) classifier, where precipitating is defined as having $f_{\text{cold}} > 10^{-2}$. \emph{Right:} The probability distribution function of temperature for varying amplitudes of compressive driving for simulations where $3.5 < t_{\rm cool}/t_{\text{ff}} < 4.5$. The amplitude of turbulent driving is shown in terms of the resulting turbulent rms Mach number.}
    \label{fig:compressive_driving}
\end{figure*}

In Figure \ref{fig:compressive_driving}, we show the outcomes of simulations with compressive driving. These are the outcomes from a parameter study of 200 simulations run with compressive driving over a range of input $t_{\rm cool}/t_{\text{ff}}$ ratio and input turbulent driving amplitude, identical to the sampling done for solenoidal driving. As before, all of the simulation properties are shown at the epoch of maximum midplane cold gas fraction. The boundary between the precipitating and non-precipitating simulations lies at higher ratios of $t_{\rm cool}/t_{\text{ff}}$, compared to solenoidal driving, for all Mach numbers. This boundary also monotonically increases with Mach number, indicating that increasing the amplitude of compressive driving makes it more likely for a simulation to become precipitating.

In Figure \ref{fig:compressive_driving} (right panel), we show the probability distribution function (PDF) of temperature for these simulations, drawn from the simulation samples within the gray box of Figure $\ref{fig:compressive_driving}$ (those having $3.5 < t_{\rm cool}/t_{\text{ff}} < 4.5$ and $\mathcal{M} < 0.5$). All of these temperature PDFs are unimodal, except for the temperature PDF corresponding to the simulation with the highest Mach number ($\mathcal{M} = 0.46$), which has an additional peak just above the temperature floor of the simulation ($10^6$ K). The other PDFs are approximately log-normal and centered near the initial temperature of $10^7$ K.

\subsubsection{Vertical driving}

\begin{figure*}
    \includegraphics[width=\textwidth]{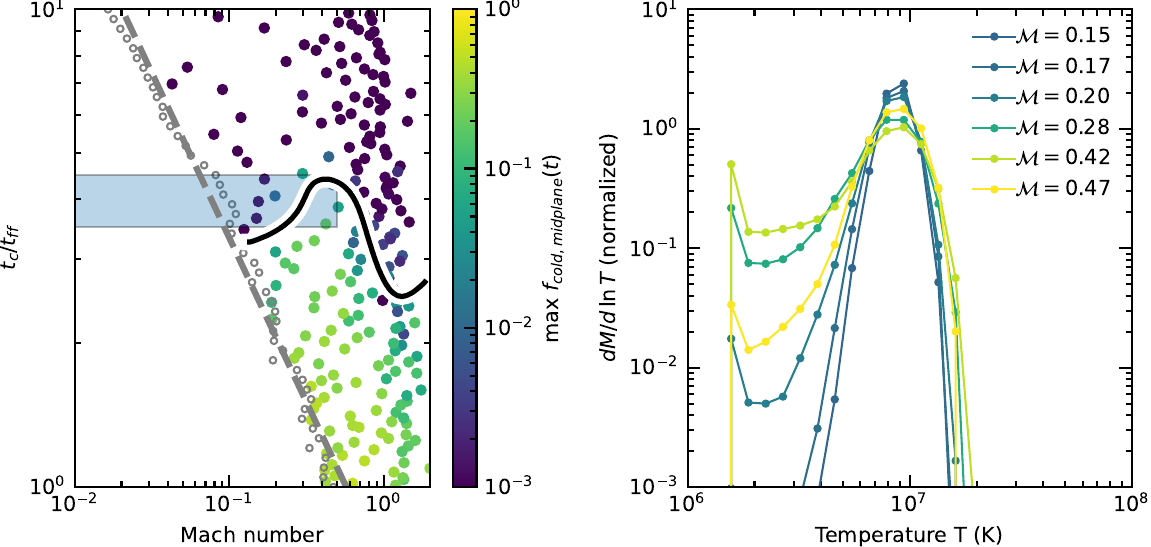}
    \caption{\emph{Left:} For vertical driving, the maximum fraction of cold gas in the midplane $f_{\text{cold}}$ as a function of cooling-to-freefall-time ratio and near-midplane rms sonic Mach number $\mathcal{M}$. The filled circles show simulations with turbulent driving, while the unfilled circles represent simulations without turbulent driving. The shaded box shows the regime from which the profiles shown in the right panel are drawn. The dashed gray line shows the best-fit power-law for the simulations without turbulent driving. The solid black line shows the boundary between precipitating and non-precipitating simulations with turbulent driving constructed using a support vector machine (SVM) classifier, where precipitating is defined as having $f_{\text{cold}} > 10^{-2}$. \emph{Right:} The probability distribution function of temperature for varying amplitudes of vertical driving for simulations where $3.5 < t_{\rm cool}/t_{\text{ff}} < 4.5$. The amplitude of turbulent driving is shown in terms of the resulting turbulent rms Mach number.}
    \label{fig:vertical_driving}
\end{figure*}

In Figure \ref{fig:vertical_driving}, the outcomes of simulations with vertical driving are shown. These are the outcomes from a parameter study of 200 simulations run with vertical driving over a range of input $t_{\rm cool}/t_{\text{ff}}$ ratio and input turbulent driving amplitude, identical to the sampling done for solenoidal and compressive driving. For this case, we observe a different boundary separating the precipitating and non-precipitating simulations. In particular, as the Mach number is increased, the boundary in $t_{\rm cool}/t_{\text{ff}}$ first increases (up to a Mach number of about 0.5) and then decreases (until a Mach number of about one). This shape means that as vertical driving is initially turned on, the ability of simulations to precipitate is increased. However, once the vertical driving causes the Mach number to exceed $\simeq 0.5$, further increases in the amplitude of vertical driving make it less likely that a simulation will precipitate at a given ratio of $t_{\rm cool}/t_{\text{ff}}$. We hypothesize that this change in behavior is due to increased mixing and shredding of cold gas clouds at higher Mach numbers that overtakes the precipitation-inducing effects of turbulent vertical transport that are produced at larger Mach numbers. Detailed investigation of this hypothesis is left to future work.

In Figure \ref{fig:vertical_driving} (right panel), we show the PDF of temperature for the simulations within the parameter space $3.5 < t_{\rm cool}/t_{\text{ff}} < 4.5$ and $\mathcal{M} < 0.5$. Unlike the previous two cases, almost all of the temperature PDFs are bimodal, with a secondary peak just above the temperature floor of the simulations. All simulations in this parameter space with Mach numbers $\mathcal{M} \ge 0.20$ have bimodal temperature PDFs. We note that the temperature PDF for the highest Mach number simulation in this parameter space ($\mathcal{M} = 0.47$) shows less cold gas than the next-highest Mach number simulation ($\mathcal{M} = 0.42$), confirming that excessively strong turbulent driving can decrease the likelihood of precipitation.

\section{Discussion}
\label{section:discussion}

The question we set out to answer when preparing this set of simulations was whether turbulence can induce precipitation in a stratified atmosphere with $t_{\rm cool}/t_{\rm ff} \approx 10$. Our results demonstrate that precipitation in such an atmosphere, if it happens, requires either modification of the background atmosphere's configuration or modification of the atmospheric conditions, including (but not limited to) addition of magnetic fields \citep[e.g.,][]{Ji_2018}. Here we discuss why some seemingly similar simulations have been able to produce precipitation in atmospheres with $t_{\rm cool}/t_{\rm ff} \approx 10$. We then compare them to our own, in order to clarify the conditions that allow circumgalactic precipitation to happen, and suggest how future work might provide further clarification.

\subsection{Precipitation in our simulations}

Our simulations confirm that driving of turbulence in a stratified galactic atmosphere with $t_{\rm cool}/t_{\rm ff} > 1$ can stimulate precipitation. They also show that vertical driving is more effective than either (isotropic) solenoidal driving or compressive driving. This is apparent from comparing the left panels of Figure \ref{fig:vertical_driving} with the left panels of Figures \ref{fig:solenoidal_driving} and \ref{fig:compressive_driving}. Vertical driving successfully induces precipitation at higher $t_{\rm cool}/t_{\text{ff}}$ ratios and lower Mach numbers, compared to the other driving modes, and it happens at Mach numbers of order $\mathcal{M} \sim 0.3$, and $t_{\rm cool}/t_{\text{ff}} \lesssim 5$. None of our driving mechanism appears to be able to induce precipitation for ratios $t_{\rm cool}/t_{\text{ff}}$ greater than 5.

For both solenoidal and vertical driving, if there is too much driving (i.e., the turbulent Mach number becomes too large), then precipitation no longer happens. We do not have a definitive understanding of why this happens, but it is consistent with the hypothesis that there is too much mixing \citep{Banerjee_2014}, so that cold clouds get quickly shredded. We emphasize that our thermostat maintains a constant temperature profile (on average), even when it requires the background average temperature rises above the thermostat value and cooling the gas is required, so it cannot be due to over-heating the medium. Curiously, the compressive driving configuration does not exhibit the phenomenon of non-monotonic cold gas fraction with Mach number. This fact is consistent with our favored hypothesis of overmixing.

Is it possible to induce more precipitation with a driving field that is more strongly correlated than a Gaussian random field in space and time? Non-Gaussian driving would be more impulsive and more `bursty,' perhaps more similar to AGN feedback as modeled in simulations and inferred from observations of X-ray cavities. We speculate that a driving mechanism with bursty, non-Gaussian correlations may allow more time for formation of cold gas blobs, because correlated bursts may lift low-entropy gas to greater altitudes, with larger values of $t_{\rm ff}$. If so, then strong bursts of AGN jet feedback may produce more precipitation than any of the idealized turbulent driving mechanisms studied in this work. We are not aware of any work that has specifically studied this question.

\subsection{Other simulations with driven turbulence}

Among recent simulations of circumgalactic precipitation in the literature, the simulations of \cite{Mohapatra_2023} are the most similar to ours. Using a suite of 16 simulations, they investigated driven turbulence in a stratified medium with radiative cooling and shell-by-shell thermal balance for a range of initial conditions. They found that compressive driving was more effective than solenoidal driving in stimulating precipitation, and we confirm that general finding. (They did not explore vertical driving.) As in our simulations, whether or not precipitation happens depends on how the median value $t_{\rm cool}/t_{\rm ff}$ compares with the dispersion in $t_{\rm cool}/t_{\rm ff}$.

In two of their simulations, \citet{Mohapatra_2023} found that compressive driving was able to stimulate precipitation in environments with $t_{\rm cool}/t_{\rm ff} \sim 10$. However, precipitation happened only after the background atmosphere has significantly evolved. In each case, the original atmosphere started out stratified, with a steadily rising entropy gradient. But by the onset of precipitation, the entropy gradient at low altitude was essentially flat, meaning that buoyancy damping could no longer happen there (see their Figure 5). Turbulent mixing causes this secular evolution in the background entropy profiles, erasing the entropy gradient over a few mixing timescales.

In contrast, we have implemented a thermostat mechanism that maintains the original entropy gradient in spite of this mixing, so that the background atmosphere always resembles the observed entropy gradients of galaxy clusters. The background entropy profile of our simulations therefore changes by $\lesssim 10\%$ from the initial conditions to their final state at a simulation elapsed time of 10 Gyr. This difference likely explains the differing boundaries we find between precipitating and non-precipitating simulations in the $t_{\rm cool}/t_{\rm ff}$--Mach number plane.

\citet{Gaspari_2013}, in the first work to show that driven turbulence could stimulate precipitation, investigated a spherical environment using a method for maintaining thermal balance similar to the one pioneered by \citet{McCourt_2012}. They found that driven turbulence with a Mach number $\mathcal{M} \sim 0.35$ was able to stimulate precipitation in environments with $t_{\rm cool}/t_{\rm ff}$ as large as $\sim 10$. However, their Figure 15 suggests that the entropy gradient of their simulated atmosphere evolves with time and may become nearly flat within the radius at which $t_{\rm cool}/t_{\rm ff} \approx 10$.

Based on such simulations, \citet{Gaspari_2018} proposed an alternative precipitation criterion. Instead of focusing on gravity and buoyancy, they focused on the competition between cooling and turbulent mixing, arguing that the key dimensionless quantity is
\begin{align}
    C \equiv t_{\text{cool}} \left( \frac{2\pi L}{\delta v} \right)^{-1} = \frac {1} {2 \pi \, \rm{Da}_t}
    \, ,
\end{align}
where $t_{\text{cool}}$ is the cooling time, $L$ is the driving scale of turbulence, and $\delta v$ is the rms velocity dispersion of the gas. For convenience, we will refer to this parameter as the \emph{cooling-mixing criterion}. Values of $C$ greater than unity inhibit precipitation, because turbulence shreds incipient cool clouds before they can precipitate, and so precipitation is expected for $C \lesssim 1$.
We note that this is equivalent to a criterion based on the turbulent Damkohler number that includes an additional factor of $2\pi$.

\begin{figure}
    \includegraphics[width=\columnwidth]{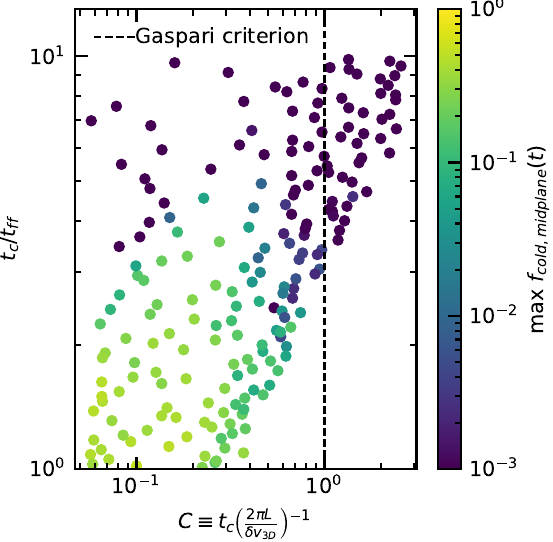}
    \caption{\emph{Left:} For vertical driving, the maximum fraction of cold gas in the midplane $f_{\text{cold}}$ as a function of cooling-to-freefall-time ratio and Gaspari precipitation parameter ${C}$. Each point represents a simulation with a different initial profile and turbulence driving amplitude.}
    \label{fig:gaspari}
\end{figure}

Figure \ref{fig:gaspari} shows how our simulation results with vertical driving relate to the \citet{Gaspari_2018} cooling-mixing criterion. We find that precipitation does not happen in our simulations with $C \gtrsim 0.8$. However, precipitation does not always happen in our simulations with $C \gtrsim 0.8$, because it is suppressed whenever $t_{\rm cool}/t_{\rm ff} \gtrsim 6$. We therefore conclude that satisfying the cooling-mixing criterion is \textit{necessary} but not \textit{sufficient} for precipitation.

\subsection{Dynamical timescales: $t_{\rm ff}$ versus $t_{\rm BV}$}

One possible explanation for why precipitation happens in some other simulations that have $t_{\rm cool}/t_{\rm ff} > 5$, but not in ours, is the difference between the free-fall timescale $t_{\rm ff}$ and the Brunt-Vaisala period $t_{\rm BV}$. When an atmosphere's entropy gradient is flat (i.e., when $d \ln Tn^{-2/3} / d \ln r \ll 1$), then the Brunt-Vaisala period and the free-fall timescale are no longer tightly connected and the Brunt-Vaisala period $t_{\text{BV}}$ becomes much longer than the free-fall timescale $t_{\text{ff}}$. In that case, buoyancy might not suppress precipitation in atmospheres with $t_{\rm cool}/t_{\rm ff} \approx 10$, for reasons discussed in \citet{Voit_2017}.

Figure 1 shows that $t_{\rm cool}/t_{\rm BV}$ in our simulations is $\approx 0.3 \, t_{\rm cool}/t_{\rm ff}$ at 5~kpc (immediately above the grey region where heating and cooling are disabled) and rises to $\approx 0.9 \, t_{\rm cool}/t_{\rm ff}$ at the atmosphere's scale height $z_0$. We note that the shape of the entropy profile is unchanged as we vary the normalization of $t_{\text{cool}}/t_{\text{ff}}$ at $z_0$, so the value of $t_{\text{cool}}/t_{\text{BV}}$ relative to $t_{\text{cool}}/t_{\text{ff}}$ as a function of height is also unchanged. Setting $t_{\rm cool}/t_{\rm ff} > 5$ at $z_0$ therefore ensures that the median value of $t_{\rm cool} / t_{\rm BV}$ exceeds unity at all altitudes where heating and cooling are not suppressed, and remains above unity because our thermostat mechanism does not allow $t_{\rm BV}$ to change significantly.

In simulations from previous work in the literature that we have discussed, the horizontally-averaged value of $t_{\rm cool} / t_{\rm BV}$ (at all altitudes) can change with time. When precipitation happens in these simulations, there are regions of the atmosphere in which the entropy gradient is much flatter than the original entropy gradient.
For the same values of $t_{\text{cool}}/t_{\text{ff}}$, it is therefore possible that buoyancy damping is \emph{not} suppressed in simulations that are allowed to develop flat entropy profiles while suppressing precipitation in our simulations.

\subsection{Precipitation in simulations with AGN feedback}

The difference between $t_{\rm ff}$ and $t_{\rm BV}$ may also be consequential in simulations designed to explore precipitation driven by more realistic AGN feedback mechanisms. The first such simulation to demonstrate how bipolar jet feedback might keep a cluster core near $t_{\rm cool}/t_{\rm ff} \approx 10$ was by \citet{Gaspari_2012ApJ...746...94G}. Many other simulations since then have  shown similar behavior \citep[e.g.,][]{LiBryan_2014ApJ...789..153L,Li_2015ApJ...811...73L,Prasad_2015ApJ...811..108P,Meece_2017ApJ...841..133M,Prasad_2018ApJ...863...62P,Nobels_2022}. Generally speaking, the galactic atmospheres in those simulations develop precipitation when cooling causes $t_{\rm cool}/t_{\rm ff}$ to drop toward unity. The ensuing rain of cold clouds fuels outbursts of central AGN feedback, resulting in atmospheric heating that raises $t_{\rm cool}/t_{\rm ff}$, and the precipitation abates when $t_{\rm cool}/t_{\rm ff}$ rises above $\sim  10$.

However, a closer look suggests that the onset of precipitation may depend more critically on $t_{\rm cool}/t_{\rm BV}$ than on $t_{\rm cool}/t_{\rm ff}$. For example, the initial round of precipitation in the simulations of \citet{LiBryan_2014ApJ...789..153L} happens when the atmosphere is homogeneous, with a significant entropy gradient. When widespread precipitation finally starts to happen, the ambient atmosphere has reached a minimum value of $t_{\rm cool}/t_{\rm ff} \approx 3$ at a distance of about $5 \, {\rm kpc}$ from the center (see their Figure 4). Likewise, precipitation occurs in the simulations of \citet{Prasad_2018ApJ...863...62P,Prasad_2020MNRAS.495..594P} shortly after the minimum value of $t_{\rm cool}/t_{\rm ff}$ drops to $\sim 3$. 

The minimum value of $t_{\rm cool}/t_{\rm ff}$ in those simulations remains small only briefly, because the precipitation rapidly triggers feedback that heats the surrounding gas. Not only does the $t_{\rm cool}/t_{\rm ff}$ ratio rapidly rise as feedback turns on, but the central entropy gradient also rapidly flattens \citep[see, e.g.,][]{Voit_2017}. Measuring a well-defined value of $t_{\rm BV}$ while that is happening is difficult because $t_{\rm BV}$ depends on the entropy gradient, which can vary chaotically in space and time, so its value can be very sensitive to the temporal and spatial scale used to compute it. However, the qualitative features of these simulations are consistent with the fact that precipitation is permitted to happen even when $t_{\rm cool}/t_{\rm ff} \sim 10$.

Analysis of a more recent AGN feedback simulation by \citet{Nobels_2022} is consistent with some of these earlier findings. Like other simulations of this type, precipitation-regulated feedback keeps the inner parts of a galaxy cluster-like environment near $\min(t_{\rm cool}/t_{\rm ff}) \sim 10$, with brief fluctuations to $\min(t_{\rm cool}/t_{\rm ff}) \lesssim 5$ and longer fluctuations to $\min(t_{\rm cool}/t_{\rm ff}) \lesssim 20$. Interestingly, though, they also looked at how AGN feedback outbursts were temporally correlated with both $t_{\rm cool}/t_{\rm ff}$ and $t_{\rm cool}/t_{\rm ff}$. They found that, on average, low values of $t_{\rm cool}/t_{\rm ff}$ \textit{preceded} those outbursts and low values of $t_{\rm cool}/t_{\rm BV}$ \textit{followed} the outbursts.
Those results are consistent with initiation of precipitation at values of $t_{\rm cool}/t_{\rm ff}$ significantly below $\sim 10$, with precipitation continuing as feedback boosts $t_{\rm cool}/t_{\rm ff}$ because it simultaneously reduces $t_{\rm cool}/t_{\rm BV}$.

\subsection{Future work}

In future work, we plan to investigate the effects of magnetic fields on precipitation. Previous work has investigated magnetic fields in simulations of stratified thermal instability without external driving of turbulence (e.g., \citealt{Ji_2018}), but there has not yet been an investigation of the effects of external turbulent driving in this context.

In this work, the Froude number is approximately constant, as we choose a constant scale height and the sonic Mach number only varies by a factor of order unity. In order to vary the Froude number by several orders of magnitude we would need to adjust the scale height accordingly. We therefore do not have a large dynamic range in Froude number, so we cannot systematically study its effect on our results. However, observed galaxy clusters have Froude numbers of order unity \citep{Wang_2023}, similar to those of our simulations, so the variation in Froude number may not be observationally relevant for the interpretation of galaxy cluster observations.

Only a single gravitational profile is investigated in this work, namely, a constant gravitational acceleration (which produces an exponential atmosphere assuming isothermality). We originally intended to investigate a power-law atmosphere (with a power-law gravitational acceleration), but numerical issues prevented this. The source of the numerical problems is that maintaining a stable numerical hydrostatic equilibrium requires the gravitational scale height to be resolved with a minimum of 10-20 cells in the vertical direction \citep{Zingale_2002}, which is not possible with a scale-free profile, as the scale height becomes arbitrarily small close to the midplane. This case is of interest because it produces a ratio $t_{\rm cool}/t_{\text{ff}}$ that is exactly constant at all heights. However, numerical investigation of this profile will require the development of numerical methods capable of simulating a hydrostatic atmosphere with an unresolved scale height. This may be possible by analytically subtracting the hydrostatic state from the equations of motion. While this will make the numerical idealization more precise this setup departs further from real galaxy cluster profiles than our current numerical setup, so it may not be useful in directly interpreting observed galaxy clusters.

The nature of external driving can make a critical difference on the outcome of precipitation, as we have shown. All of the turbulent driving schemes we have explored depend on the idealization of a Gaussian random field, but the forces applied by AGN feedback are likely significantly more impulsive and more spatially correlated than Gaussian random fields. Investigating precipitation with a driving mechanism designed to be more similar to that of AGN jet feedback is therefore a high priority for future work.

Finally, all of our simulations were vertically stratified in a Cartesian geometry, rather than radially stratified in a spherical geometry. The spherical stratification of real galaxy cluster atmospheres may have a significant impact on precipitation outcomes due to geometric factors that are not captured in our simulations, and their impact is a potential explanation for the tension between the results of \cite{Gaspari_2013} and this work.

\section{Conclusion}
\label{section:conclusion}

We have carried out the most extensive suite of simulations to date (more than 600 simulations evolved for 10 Gyr each) aimed at investigating the precipitation hypothesis of self-regulation in galaxy clusters and the circumgalactic medium. We find qualitative agreement with the precipitation hypothesis \citep[e.g.,][]{Donahue_2022}, but the maximum value of cooling time to free-fall time ratio for precipitation found in our simulations is $t_{\rm cool}/t_{\text{ff}} \sim 5$, rather than a value of 10 that is suggested by observations of both galaxy clusters and massive elliptical galaxies \citep{Voit_2015Natur.519..203V}, and as observed in self-regulating AGN feedback simulations \citep[e.g.][]{Gaspari_2012ApJ...746...94G,Li_2015ApJ...811...73L,Nobels_2022}. To summarize, we have found that:
\begin{itemize}
    \item Our simulations maintain their initial background entropy profile as they evolve (unlike previous simulations in the literature), thanks to the use of a `thermostat' mechanism (as explained in section \ref{section:methods});
    \item The mode of extrinsic turbulent driving significant alters the parameter regime where precipitation occurs in simulations, confirming the results of \cite{Mohapatra_2023};
    \item Even more so than compressive driving, purely-vertical turbulent driving produces precipitation across the largest range of parameter space;
    \item The precipitation criterion $C$ proposed by \cite{Gaspari_2018}, which is inversely proportional the atmosphere's Damkohler number $\text{Da}_t$, is a \emph{necessary} but not \emph{sufficient} condition for precipitation to occur in simulations;
    \item Precipitation occurs in our idealized simulations for $t_{\text{cool}}/t_{\text{ff}} \lesssim 5$, in tension with observational results that suggest that precipitation should occur whenever $t_{\text{cool}}/t_{\text{ff}} \lesssim 10$.
\end{itemize}

Future work should investigate the effects of magnetic fields, spherical gravitational stratification, and impulsive feedback in order to determine whether there are any significant missing ingredients in our numerical models of precipitation that may change the critical threshold of $t_{\rm cool}/t_{\text{ff}}$ and bridge the gap between our idealized numerical experiments and the greater realism of self-regulated AGN feedback simulations. 

\section*{Acknowledgements}
We thank Philip Grete and Forrest Glines for technical assistance with the use of the AthenaPK MHD code.  The authors acknowledge the support of NSF grant \#AAG-2106575, which is the source of support for B.D.W.

B.D.W. thanks Michael Zingale for suggesting the use of well-balanced reconstruction methods and reflecting vertical boundary conditions for hydrodynamics simulations near hydrostatic equilibrium.

BWO acknowledges support from NSF grant \#AAG-1908109, NASA ATP grants NNX15AP39G and 80NSSC18K1105, and NASA TCAN grant 80NSSC21K1053. 

The simulations were carried out in part using ACCESS resources using allocation MCA08X028 (TG-AST090040; PI: Brian O'Shea). This work was also supported in part through computational resources and services provided by the Institute for Cyber-Enabled Research at Michigan State University.

\section*{Data Availability}

The simulation data underlying this article are available upon request from the authors.



\bibliographystyle{mnras}
\bibliography{precipitator}

\begin{thebibliography}{}
\makeatletter
\relax
\def\mn@urlcharsother{\let\do\@makeother \do\$\do\&\do\#\do\^\do\_\do\%\do\~}
\def\mn@doi{\begingroup\mn@urlcharsother \@ifnextchar [ {\mn@doi@}
  {\mn@doi@[]}}
\def\mn@doi@[#1]#2{\def\@tempa{#1}\ifx\@tempa\@empty \href
  {http://dx.doi.org/#2} {doi:#2}\else \href {http://dx.doi.org/#2} {#1}\fi
  \endgroup}
\def\mn@eprint#1#2{\mn@eprint@#1:#2::\@nil}
\def\mn@eprint@arXiv#1{\href {http://arxiv.org/abs/#1} {{\tt arXiv:#1}}}
\def\mn@eprint@dblp#1{\href {http://dblp.uni-trier.de/rec/bibtex/#1.xml}
  {dblp:#1}}
\def\mn@eprint@#1:#2:#3:#4\@nil{\def\@tempa {#1}\def\@tempb {#2}\def\@tempc
  {#3}\ifx \@tempc \@empty \let \@tempc \@tempb \let \@tempb \@tempa \fi \ifx
  \@tempb \@empty \def\@tempb {arXiv}\fi \@ifundefined
  {mn@eprint@\@tempb}{\@tempb:\@tempc}{\expandafter \expandafter \csname
  mn@eprint@\@tempb\endcsname \expandafter{\@tempc}}}

\bibitem[\protect\citeauthoryear{{Balbus} \& {Soker}}{{Balbus} \&
  {Soker}}{1989}]{Balbus_1989}
{Balbus} S.~A.,  {Soker} N.,  1989, \mn@doi [\apj] {10.1086/167521}, \href
  {https://ui.adsabs.harvard.edu/abs/1989ApJ...341..611B} {341, 611}

\bibitem[\protect\citeauthoryear{{Banerjee} \& {Sharma}}{{Banerjee} \&
  {Sharma}}{2014}]{Banerjee_2014}
{Banerjee} N.,  {Sharma} P.,  2014, \mn@doi [\mnras] {10.1093/mnras/stu1179},
  \href {https://ui.adsabs.harvard.edu/abs/2014MNRAS.443..687B} {443, 687}

\bibitem[\protect\citeauthoryear{Berblinger \& Schlier}{Berblinger \&
  Schlier}{1991}]{Berblinger_1991}
Berblinger M.,  Schlier C.,  1991, \mn@doi [Computer Physics Communications]
  {https://doi.org/10.1016/0010-4655(91)90064-R}, 66, 157

\bibitem[\protect\citeauthoryear{{Binney}, {Nipoti}  \& {Fraternali}}{{Binney}
  et~al.}{2009}]{Binney_2009MNRAS.397.1804B}
{Binney} J.,  {Nipoti} C.,   {Fraternali} F.,  2009, \mn@doi [\mnras]
  {10.1111/j.1365-2966.2009.15113.x}, \href
  {https://ui.adsabs.harvard.edu/abs/2009MNRAS.397.1804B} {397, 1804}

\bibitem[\protect\citeauthoryear{{Cavagnolo}, {Donahue}, {Voit}  \&
  {Sun}}{{Cavagnolo} et~al.}{2009}]{Cavagnolo_2009}
{Cavagnolo} K.~W.,  {Donahue} M.,  {Voit} G.~M.,   {Sun} M.,  2009, \mn@doi
  [\apjs] {10.1088/0067-0049/182/1/12}, \href
  {https://ui.adsabs.harvard.edu/abs/2009ApJS..182...12C} {182, 12}

\bibitem[\protect\citeauthoryear{{Choudhury} \& {Sharma}}{{Choudhury} \&
  {Sharma}}{2016}]{Choudhury_2016}
{Choudhury} P.~P.,  {Sharma} P.,  2016, \mn@doi [\mnras]
  {10.1093/mnras/stw152}, \href
  {https://ui.adsabs.harvard.edu/abs/2016MNRAS.457.2554C} {457, 2554}

\bibitem[\protect\citeauthoryear{{Choudhury}, {Kauffmann}  \&
  {Sharma}}{{Choudhury} et~al.}{2019a}]{Choudhury_2019a}
{Choudhury} P.~P.,  {Kauffmann} G.,   {Sharma} P.,  2019a, \mn@doi [\mnras]
  {10.1093/mnras/stz567}, \href
  {https://ui.adsabs.harvard.edu/abs/2019MNRAS.485.3430C} {485, 3430}

\bibitem[\protect\citeauthoryear{{Choudhury}, {Sharma}  \&
  {Quataert}}{{Choudhury} et~al.}{2019b}]{Choudhury_2019b}
{Choudhury} P.~P.,  {Sharma} P.,   {Quataert} E.,  2019b, \mn@doi [\mnras]
  {10.1093/mnras/stz1857}, \href
  {https://ui.adsabs.harvard.edu/abs/2019MNRAS.488.3195C} {488, 3195}

\bibitem[\protect\citeauthoryear{{Colella} \& {Woodward}}{{Colella} \&
  {Woodward}}{1984}]{Colella_1984}
{Colella} P.,  {Woodward} P.~R.,  1984, \mn@doi [Journal of Computational
  Physics] {10.1016/0021-9991(84)90143-8}, \href
  {https://ui.adsabs.harvard.edu/abs/1984JCoPh..54..174C} {54, 174}

\bibitem[\protect\citeauthoryear{{Cowie}, {Fabian}  \& {Nulsen}}{{Cowie}
  et~al.}{1980}]{Cowie_1980}
{Cowie} L.~L.,  {Fabian} A.~C.,   {Nulsen} P.~E.~J.,  1980, \mn@doi [\mnras]
  {10.1093/mnras/191.2.399}, \href
  {https://ui.adsabs.harvard.edu/abs/1980MNRAS.191..399C} {191, 399}

\bibitem[\protect\citeauthoryear{{Defouw}}{{Defouw}}{1970}]{Defouw_1970}
{Defouw} R.~J.,  1970, \mn@doi [\apj] {10.1086/150460}, \href
  {https://ui.adsabs.harvard.edu/abs/1970ApJ...160..659D} {160, 659}

\bibitem[\protect\citeauthoryear{{Donahue} \& {Voit}}{{Donahue} \&
  {Voit}}{2022}]{Donahue_2022}
{Donahue} M.,  {Voit} G.~M.,  2022, \mn@doi [\physrep]
  {10.1016/j.physrep.2022.04.005}, \href
  {https://ui.adsabs.harvard.edu/abs/2022PhR...973....1D} {973, 1}

\bibitem[\protect\citeauthoryear{{Eswaran} \& {Pope}}{{Eswaran} \&
  {Pope}}{1988}]{Eswaran_1988}
{Eswaran} V.,  {Pope} S.~B.,  1988, Computers and Fluids, \href
  {https://ui.adsabs.harvard.edu/abs/1988CF.....16..257E} {16, 257}

\bibitem[\protect\citeauthoryear{Feinberg \& Langtangen}{Feinberg \&
  Langtangen}{2015}]{Feinberg_2015}
Feinberg J.,  Langtangen H.~P.,  2015, \mn@doi [Journal of Computational
  Science] {https://doi.org/10.1016/j.jocs.2015.08.008}, 11, 46

\bibitem[\protect\citeauthoryear{{Field}}{{Field}}{1965}]{Field_1965}
{Field} G.~B.,  1965, \mn@doi [\apj] {10.1086/148317}, \href
  {https://ui.adsabs.harvard.edu/abs/1965ApJ...142..531F} {142, 531}

\bibitem[\protect\citeauthoryear{{Gaspari}, {Ruszkowski}  \&
  {Sharma}}{{Gaspari} et~al.}{2012}]{Gaspari_2012ApJ...746...94G}
{Gaspari} M.,  {Ruszkowski} M.,   {Sharma} P.,  2012, \mn@doi [\apj]
  {10.1088/0004-637X/746/1/94}, \href
  {https://ui.adsabs.harvard.edu/abs/2012ApJ...746...94G} {746, 94}

\bibitem[\protect\citeauthoryear{{Gaspari}, {Ruszkowski}  \& {Oh}}{{Gaspari}
  et~al.}{2013}]{Gaspari_2013}
{Gaspari} M.,  {Ruszkowski} M.,   {Oh} S.~P.,  2013, \mn@doi [\mnras]
  {10.1093/mnras/stt692}, \href
  {https://ui.adsabs.harvard.edu/abs/2013MNRAS.432.3401G} {432, 3401}

\bibitem[\protect\citeauthoryear{{Gaspari} et~al.,}{{Gaspari}
  et~al.}{2018}]{Gaspari_2018}
{Gaspari} M.,  et~al., 2018, \mn@doi [\apj] {10.3847/1538-4357/aaaa1b}, \href
  {https://ui.adsabs.harvard.edu/abs/2018ApJ...854..167G} {854, 167}

\bibitem[\protect\citeauthoryear{{Grete}, {O'Shea}  \& {Beckwith}}{{Grete}
  et~al.}{2018}]{Grete_2018}
{Grete} P.,  {O'Shea} B.~W.,   {Beckwith} K.,  2018, \mn@doi [\apjl]
  {10.3847/2041-8213/aac0f5}, \href
  {https://ui.adsabs.harvard.edu/abs/2018ApJ...858L..19G} {858, L19}

\bibitem[\protect\citeauthoryear{{Grete} et~al.,}{{Grete}
  et~al.}{2022}]{Grete_2022}
{Grete} P.,  et~al., 2022, \mn@doi [arXiv e-prints]
  {10.48550/arXiv.2202.12309}, \href
  {https://ui.adsabs.harvard.edu/abs/2022arXiv220212309G} {p. arXiv:2202.12309}

\bibitem[\protect\citeauthoryear{Guillard \& Murrone}{Guillard \&
  Murrone}{2004}]{Guillard_2004}
Guillard H.,  Murrone A.,  2004, \mn@doi [Computers & Fluids]
  {https://doi.org/10.1016/j.compfluid.2003.07.001}, 33, 655

\bibitem[\protect\citeauthoryear{Halton}{Halton}{1960}]{Halton_1960}
Halton J.~H.,  1960, \mn@doi [Numerische Mathematik] {10.1007/BF01386213}, 2,
  84

\bibitem[\protect\citeauthoryear{{Hitomi Collaboration} et~al.,}{{Hitomi
  Collaboration} et~al.}{2016}]{Hitomi_2016}
{Hitomi Collaboration} et~al., 2016, \mn@doi [\nat] {10.1038/nature18627},
  \href {https://ui.adsabs.harvard.edu/abs/2016Natur.535..117H} {535, 117}

\bibitem[\protect\citeauthoryear{{Hitomi Collaboration} et~al.,}{{Hitomi
  Collaboration} et~al.}{2018}]{Hitomi_2018}
{Hitomi Collaboration} et~al., 2018, \mn@doi [\pasj] {10.1093/pasj/psx138},
  \href {https://ui.adsabs.harvard.edu/abs/2018PASJ...70....9H} {70, 9}

\bibitem[\protect\citeauthoryear{{Ji}, {Oh}  \& {McCourt}}{{Ji}
  et~al.}{2018}]{Ji_2018}
{Ji} S.,  {Oh} S.~P.,   {McCourt} M.,  2018, \mn@doi [\mnras]
  {10.1093/mnras/sty293}, \href
  {https://ui.adsabs.harvard.edu/abs/2018MNRAS.476..852J} {476, 852}

\bibitem[\protect\citeauthoryear{{K{\"a}ppeli} \& {Mishra}}{{K{\"a}ppeli} \&
  {Mishra}}{2014}]{Kappeli_2014}
{K{\"a}ppeli} R.,  {Mishra} S.,  2014, \mn@doi [Journal of Computational
  Physics] {10.1016/j.jcp.2013.11.028}, \href
  {https://ui.adsabs.harvard.edu/abs/2014JCoPh.259..199K} {259, 199}

\bibitem[\protect\citeauthoryear{{Li} \& {Bryan}}{{Li} \&
  {Bryan}}{2014}]{LiBryan_2014ApJ...789..153L}
{Li} Y.,  {Bryan} G.~L.,  2014, \mn@doi [\apj] {10.1088/0004-637X/789/2/153},
  \href {https://ui.adsabs.harvard.edu/abs/2014ApJ...789..153L} {789, 153}

\bibitem[\protect\citeauthoryear{{Li}, {Bryan}, {Ruszkowski}, {Voit}, {O'Shea}
  \& {Donahue}}{{Li} et~al.}{2015}]{Li_2015ApJ...811...73L}
{Li} Y.,  {Bryan} G.~L.,  {Ruszkowski} M.,  {Voit} G.~M.,  {O'Shea} B.~W.,
  {Donahue} M.,  2015, \mn@doi [\apj] {10.1088/0004-637X/811/2/73}, \href
  {https://ui.adsabs.harvard.edu/abs/2015ApJ...811...73L} {811, 73}

\bibitem[\protect\citeauthoryear{McCorquodale \& Colella}{McCorquodale \&
  Colella}{2011}]{McCorquodale_2011}
McCorquodale P.,  Colella P.,  2011, \mn@doi [Communications in Applied
  Mathematics and Computational Science] {10.2140/camcos.2011.6.1}, 6, 1

\bibitem[\protect\citeauthoryear{{McCourt}, {Sharma}, {Quataert}  \&
  {Parrish}}{{McCourt} et~al.}{2012}]{McCourt_2012}
{McCourt} M.,  {Sharma} P.,  {Quataert} E.,   {Parrish} I.~J.,  2012, \mn@doi
  [\mnras] {10.1111/j.1365-2966.2011.19972.x}, \href
  {https://ui.adsabs.harvard.edu/abs/2012MNRAS.419.3319M} {419, 3319}

\bibitem[\protect\citeauthoryear{{McNamara}, {Russell}, {Nulsen}, {Hogan},
  {Fabian}, {Pulido}  \& {Edge}}{{McNamara} et~al.}{2016}]{McNamara_2016}
{McNamara} B.~R.,  {Russell} H.~R.,  {Nulsen} P.~E.~J.,  {Hogan} M.~T.,
  {Fabian} A.~C.,  {Pulido} F.,   {Edge} A.~C.,  2016, \mn@doi [\apj]
  {10.3847/0004-637X/830/2/79}, \href
  {https://ui.adsabs.harvard.edu/abs/2016ApJ...830...79M} {830, 79}

\bibitem[\protect\citeauthoryear{{Meece}, {O'Shea}  \& {Voit}}{{Meece}
  et~al.}{2015}]{Meece_2015}
{Meece} G.~R.,  {O'Shea} B.~W.,   {Voit} G.~M.,  2015, \mn@doi [\apj]
  {10.1088/0004-637X/808/1/43}, \href
  {https://ui.adsabs.harvard.edu/abs/2015ApJ...808...43M} {808, 43}

\bibitem[\protect\citeauthoryear{{Meece}, {Voit}  \& {O'Shea}}{{Meece}
  et~al.}{2017}]{Meece_2017ApJ...841..133M}
{Meece} G.~R.,  {Voit} G.~M.,   {O'Shea} B.~W.,  2017, \mn@doi [\apj]
  {10.3847/1538-4357/aa6fb1}, \href
  {https://ui.adsabs.harvard.edu/abs/2017ApJ...841..133M} {841, 133}

\bibitem[\protect\citeauthoryear{{Minoshima} \& {Miyoshi}}{{Minoshima} \&
  {Miyoshi}}{2021}]{Minoshima_2021}
{Minoshima} T.,  {Miyoshi} T.,  2021, \mn@doi [Journal of Computational
  Physics] {10.1016/j.jcp.2021.110639}, \href
  {https://ui.adsabs.harvard.edu/abs/2021JCoPh.44610639M} {446, 110639}

\bibitem[\protect\citeauthoryear{{Mohapatra} \& {Quataert}}{{Mohapatra} \&
  {Quataert}}{2024}]{Mohapatra_2024}
{Mohapatra} R.,  {Quataert} E.,  2024, \mn@doi [\apj]
  {10.3847/1538-4357/ad2940}, \href
  {https://ui.adsabs.harvard.edu/abs/2024ApJ...965..105M} {965, 105}

\bibitem[\protect\citeauthoryear{{Mohapatra} \& {Sharma}}{{Mohapatra} \&
  {Sharma}}{2019}]{Mohapatra_2019}
{Mohapatra} R.,  {Sharma} P.,  2019, \mn@doi [\mnras] {10.1093/mnras/stz328},
  \href {https://ui.adsabs.harvard.edu/abs/2019MNRAS.484.4881M} {484, 4881}

\bibitem[\protect\citeauthoryear{{Mohapatra}, {Jetti}, {Sharma}  \&
  {Federrath}}{{Mohapatra} et~al.}{2022}]{Mohapatra_2022}
{Mohapatra} R.,  {Jetti} M.,  {Sharma} P.,   {Federrath} C.,  2022, \mn@doi
  [\mnras] {10.1093/mnras/stab3603}, \href
  {https://ui.adsabs.harvard.edu/abs/2022MNRAS.510.3778M} {510, 3778}

\bibitem[\protect\citeauthoryear{{Mohapatra}, {Sharma}, {Federrath}  \&
  {Quataert}}{{Mohapatra} et~al.}{2023}]{Mohapatra_2023}
{Mohapatra} R.,  {Sharma} P.,  {Federrath} C.,   {Quataert} E.,  2023, \mn@doi
  [\mnras] {10.1093/mnras/stad2574}, \href
  {https://ui.adsabs.harvard.edu/abs/2023MNRAS.525.3831M} {525, 3831}

\bibitem[\protect\citeauthoryear{{Nobels}, {Schaye}, {Schaller}, {Bah{\'e}}  \&
  {Chaikin}}{{Nobels} et~al.}{2022}]{Nobels_2022}
{Nobels} F. S.~J.,  {Schaye} J.,  {Schaller} M.,  {Bah{\'e}} Y.~M.,   {Chaikin}
  E.,  2022, \mn@doi [\mnras] {10.1093/mnras/stac2061}, \href
  {https://ui.adsabs.harvard.edu/abs/2022MNRAS.515.4838N} {515, 4838}

\bibitem[\protect\citeauthoryear{{Nulsen}}{{Nulsen}}{1986}]{Nulsen_1986}
{Nulsen} P.~E.~J.,  1986, \mn@doi [\mnras] {10.1093/mnras/221.2.377}, \href
  {https://ui.adsabs.harvard.edu/abs/1986MNRAS.221..377N} {221, 377}

\bibitem[\protect\citeauthoryear{{Prasad}, {Sharma}  \& {Babul}}{{Prasad}
  et~al.}{2015}]{Prasad_2015ApJ...811..108P}
{Prasad} D.,  {Sharma} P.,   {Babul} A.,  2015, \mn@doi [\apj]
  {10.1088/0004-637X/811/2/108}, \href
  {https://ui.adsabs.harvard.edu/abs/2015ApJ...811..108P} {811, 108}

\bibitem[\protect\citeauthoryear{{Prasad}, {Sharma}  \& {Babul}}{{Prasad}
  et~al.}{2018}]{Prasad_2018ApJ...863...62P}
{Prasad} D.,  {Sharma} P.,   {Babul} A.,  2018, \mn@doi [\apj]
  {10.3847/1538-4357/aacce8}, \href
  {https://ui.adsabs.harvard.edu/abs/2018ApJ...863...62P} {863, 62}

\bibitem[\protect\citeauthoryear{{Prasad}, {Sharma}, {Babul}, {Voit}  \&
  {O'Shea}}{{Prasad} et~al.}{2020}]{Prasad_2020MNRAS.495..594P}
{Prasad} D.,  {Sharma} P.,  {Babul} A.,  {Voit} G.~M.,   {O'Shea} B.~W.,  2020,
  \mn@doi [\mnras] {10.1093/mnras/staa1247}, \href
  {https://ui.adsabs.harvard.edu/abs/2020MNRAS.495..594P} {495, 594}

\bibitem[\protect\citeauthoryear{{Schwarzschild}}{{Schwarzschild}}{1958}]{Schwarzschild_1958ses..book.....S}
{Schwarzschild} M.,  1958, {Structure and evolution of the stars.}

\bibitem[\protect\citeauthoryear{{Sharma}, {McCourt}, {Quataert}  \&
  {Parrish}}{{Sharma} et~al.}{2012}]{Sharma_2012}
{Sharma} P.,  {McCourt} M.,  {Quataert} E.,   {Parrish} I.~J.,  2012, \mn@doi
  [\mnras] {10.1111/j.1365-2966.2011.20246.x}, \href
  {https://ui.adsabs.harvard.edu/abs/2012MNRAS.420.3174S} {420, 3174}

\bibitem[\protect\citeauthoryear{{Shu} \& {Osher}}{{Shu} \&
  {Osher}}{1989}]{Shu_1989}
{Shu} C.-W.,  {Osher} S.,  1989, \mn@doi [Journal of Computational Physics]
  {10.1016/0021-9991(89)90222-2}, \href
  {https://ui.adsabs.harvard.edu/abs/1989JCoPh..83...32S} {83, 32}

\bibitem[\protect\citeauthoryear{{Spitzer}}{{Spitzer}}{1956}]{Spitzer_1956}
{Spitzer} Lyman J.,  1956, \mn@doi [\apj] {10.1086/146200}, \href
  {https://ui.adsabs.harvard.edu/abs/1956ApJ...124...20S} {124, 20}

\bibitem[\protect\citeauthoryear{Thornber, Drikakis, Williams  \&
  Youngs}{Thornber et~al.}{2008}]{Thornber_2008}
Thornber B.,  Drikakis D.,  Williams R.,   Youngs D.,  2008, \mn@doi [Journal
  of Computational Physics] {https://doi.org/10.1016/j.jcp.2008.01.035}, 227,
  4853

\bibitem[\protect\citeauthoryear{{Voit}}{{Voit}}{2018}]{Voit_2018}
{Voit} G.~M.,  2018, \mn@doi [\apj] {10.3847/1538-4357/aae8e2}, \href
  {https://ui.adsabs.harvard.edu/abs/2018ApJ...868..102V} {868, 102}

\bibitem[\protect\citeauthoryear{{Voit}}{{Voit}}{2021}]{Voit_2021}
{Voit} G.~M.,  2021, \mn@doi [\apjl] {10.3847/2041-8213/abe11f}, \href
  {https://ui.adsabs.harvard.edu/abs/2021ApJ...908L..16V} {908, L16}

\bibitem[\protect\citeauthoryear{{Voit}, {Donahue}, {Bryan}  \&
  {McDonald}}{{Voit} et~al.}{2015}]{Voit_2015Natur.519..203V}
{Voit} G.~M.,  {Donahue} M.,  {Bryan} G.~L.,   {McDonald} M.,  2015, \mn@doi
  [\nat] {10.1038/nature14167}, \href
  {https://ui.adsabs.harvard.edu/abs/2015Natur.519..203V} {519, 203}

\bibitem[\protect\citeauthoryear{{Voit}, {Meece}, {Li}, {O'Shea}, {Bryan}  \&
  {Donahue}}{{Voit} et~al.}{2017}]{Voit_2017}
{Voit} G.~M.,  {Meece} G.,  {Li} Y.,  {O'Shea} B.~W.,  {Bryan} G.~L.,
  {Donahue} M.,  2017, \mn@doi [\apj] {10.3847/1538-4357/aa7d04}, \href
  {https://ui.adsabs.harvard.edu/abs/2017ApJ...845...80V} {845, 80}

\bibitem[\protect\citeauthoryear{{Wang}, {Oh}  \& {Ruszkowski}}{{Wang}
  et~al.}{2023}]{Wang_2023}
{Wang} C.,  {Oh} S.~P.,   {Ruszkowski} M.,  2023, \mn@doi [\mnras]
  {10.1093/mnras/stad003}, \href
  {https://ui.adsabs.harvard.edu/abs/2023MNRAS.519.4408W} {519, 4408}

\bibitem[\protect\citeauthoryear{{Zhuravleva} et~al.,}{{Zhuravleva}
  et~al.}{2015}]{Zhuravleva_2015}
{Zhuravleva} I.,  et~al., 2015, \mn@doi [\mnras] {10.1093/mnras/stv900}, \href
  {https://ui.adsabs.harvard.edu/abs/2015MNRAS.450.4184Z} {450, 4184}

\bibitem[\protect\citeauthoryear{{Zhuravleva}, {Chen}, {Churazov},
  {Schekochihin}, {Zhang}  \& {Nagai}}{{Zhuravleva}
  et~al.}{2023}]{Zhuravleva_2023}
{Zhuravleva} I.,  {Chen} M.~C.,  {Churazov} E.,  {Schekochihin} A.~A.,  {Zhang}
  C.,   {Nagai} D.,  2023, \mn@doi [\mnras] {10.1093/mnras/stad470}, \href
  {https://ui.adsabs.harvard.edu/abs/2023MNRAS.520.5157Z} {520, 5157}

\bibitem[\protect\citeauthoryear{{Zingale} et~al.,}{{Zingale}
  et~al.}{2002}]{Zingale_2002}
{Zingale} M.,  et~al., 2002, \mn@doi [\apjs] {10.1086/342754}, \href
  {https://ui.adsabs.harvard.edu/abs/2002ApJS..143..539Z} {143, 539}

\makeatother
\end{thebibliography}






\bsp	
\label{lastpage}
\end{document}